\documentclass[superscriptaddress,nofootinbib,twocolumn,letterpaper,prd]{revtex4-1}

\usepackage{amsmath,amssymb}
\usepackage{graphicx}
\usepackage[usenames,dvipsnames]{xcolor}
\usepackage{hyperref}
\usepackage{url}
\usepackage[utf8]{inputenc}
\usepackage{slashed}
\usepackage{subfigure}
\usepackage{slashed,bbm}
\usepackage{graphicx,psfrag,epsfig}
\usepackage{dsfont}
\usepackage{setspace}
\usepackage{blindtext}
\usepackage{enumerate}
\hypersetup{
    colorlinks,
    linkcolor={blue},
    citecolor={blue},
    urlcolor={blue}
}

\newcommand{\cpone}{CP${}^1$}
\newcommand{\cpn}{CP${}^{N_b - 1}$}
\newcommand{\qedgn}{QED$_3$-GN}
\newcommand{\Nf}{N}
\newcommand{\Nb}{N_\mathrm{b}}

\DeclareMathOperator{\Real}{Re}
\DeclareMathOperator{\Imag}{Im}

\allowdisplaybreaks[1]

\begin{document}

\title{Deconfined criticality from the QED$_3$-Gross-Neveu model at three loops}

\author{Bernhard Ihrig}
\affiliation{Institute for Theoretical Physics, University of Cologne, Z\"ulpicher Stra\ss e 77, 50937 Cologne, Germany}

\author{Lukas Janssen}
\affiliation{Institut f\"ur Theoretische Physik, Technische Universit\"at Dresden, 01062 Dresden, Germany}

\author{Luminita~N.~Mihaila}
\affiliation{Institute for Theoretical Physics, Heidelberg University, Philosophenweg 16, 69120 Heidelberg, Germany}

\author{Michael~M.~Scherer}
\affiliation{Institute for Theoretical Physics, University of Cologne, Z\"ulpicher Stra\ss e 77, 50937 Cologne, Germany}

\begin{abstract}

The QED$_3$-Gross-Neveu model is a $(2+1)$-dimensional $\mathrm U(1)$ gauge theory involving Dirac fermions and a critical real scalar field. This theory has recently been argued to represent a dual description of the deconfined quantum critical point between N\'eel and valence bond solid orders in frustrated quantum magnets. We study the critical behavior of the QED$_3$-Gross-Neveu model by means of an $\epsilon$ expansion around the upper critical space-time dimension of $D_\mathrm{c}^+=4$ up to the three-loop order.
Estimates for critical exponents in $2+1$ dimensions are obtained by evaluating the different Pad\'e approximants of their series expansion in $\epsilon$.
We find that these estimates, within the spread of the Pad\'e approximants, satisfy a nontrivial scaling relation which follows from the emergent $\mathrm{SO}(5)$~symmetry implied by the duality conjecture.
We also construct explicit evidence for the equivalence between the QED$_3$-Gross-Neveu model and a corresponding critical four-fermion gauge theory that was previously studied within the $1/N$ expansion in space-time dimensions $2<D<4$.

\end{abstract}

\maketitle

\section{Introduction}

Frustrated magnets host a variety of nontrivial quantum and classical states in their low-energy spectrum. Upon changing nonthermal external parameters such as magnetic field, pressure, or chemical composition these systems, in the low-temperature limit, can often be tuned through quantum critical points (QCPs) at which the nature of the ground state changes qualitatively \cite{sachdevbook}.
Apart from the featureless disordered states, we distinguish between long-range-ordered ground states that are characterized by local order parameters and describe symmetry-broken phases, and unconventional long-range-entangled states that exhibit fractionalized excitations.
In a Heisenberg magnet on the square lattice, for example, the classical N\'eel state corresponding to the standard antiferromagnetic phase spontaneously breaks the spin-rotational symmetry. The valence bond solid (VBS), in which pairs of spins on neighboring sites form singlets, is another example for a conventionally-ordered, yet quantum, state. It breaks the lattice-rotational symmetry spontaneously.
Unconventional long-range-entangled states, i.e., spin-liquid states, can be stabilized in situations in which the magnetic frustration is large. Their non-local fractionalized excitations lead to fascinating new physics~\cite{savary2017}.

Such unconventional behavior can, however, also occur right at the transition point between two completely conventionally-ordered phases. The prototype example for such an exotic transition is the putative quantum critical point between N\'eel and VBS states in square-lattice Heisenberg magnets~\cite{senthil2004, nahum2015a, shao2016}, but evidence for various similar critical points between two ordered phases in other systems has been found recently~\cite{sato2017, qin2017, li2017, zhang2018}.
It has been argued~\cite{senthil2004} that the low-temperature physics in the quantum critical fan~\cite{melko2008} is governed by fractionalized particles (``deconfined spinons'') which represent the relevant physical degrees of freedom at the QCP, but are ``confined'' in either of the adjacent ordered phases.
The spinons interact via an emergent gauge field, and a possible low-energy field theory describing the deconfined QCP between N\'eel and VBS states is given by the $(2+1)$-dimensional (2+1D) noncompact \cpone\ sigma model~\cite{senthil2004}.

There exist, however, other equivalent continuum descriptions of deconfined QCPs. The N\'eel-VBS deconfined critical point, for instance, possesses an alternate formulation in terms of its original order parameters provided that a certain topological term is included in the action~\cite{senthil2006}. The presence of this additional term naturally explains why the conventional Ginzburg-Landau-Wilson paradigm, which would forbid a direct and continuous transition between two ordered phases, breaks down at a deconfined QCP.
Yet another equivalent description was conjectured, which involves critical fermion degrees of freedom~\cite{wang2017}. Early indications for such a 2+1D boson-fermion duality at the critical point have been found some time ago, suggesting the equivalence between 2+1D quantum electrodynamics (QED$_3$) and an easy-plane version of the \cpone\ sigma model at criticality~\cite{alicea2005, senthil2006}.
Initiated partly by the Dirac theory of the half-filled Landau level~\cite{son2015} and the fermionic counterpart~\cite{wang2015, metlitski2016, mross2016} of the classic bosonic particle-vortex duality~\cite{peskin1978, dasgupta1981}, these early conjectures have recently been significantly extended~\cite{seiberg2016} and put into context~\cite{karch2016}. There is now a comprehensive web of dualities between 2+1D field theories including both fermionic as well as purely bosonic gauge theories.
Some of these dualities can be explicitly derived within a lattice formulation~\cite{chen2018}, others follow perturbing established supersymmetric dualities~\cite{kachru2017}, or can be verified in the large-$N$ limit~\cite{aharony2016}.
Numerical evidence for the proposed duality between the easy-plane \cpone\ sigma model and QED$_3$ has been purported very recently as well~\cite{qin2017, zhang2018}.

The $\mathrm{SU}(2)$ symmetric noncompact \cpone\ model, describing the N\'eel-VBS deconfined QCP, has been conjectured to be dual to the QED$_3$-Gross-Neveu (\qedgn) model~\cite{wang2017}. This latter theory consists of 2+1D gapless Dirac fermions that are charged under a $\mathrm U(1)$ gauge field as in QED$_3$, but are additionally coupled to a critical Gross-Neveu scalar field.
The duality implies an emergent $\mathrm{SO}(5)$ symmetry at the deconfined QCP~\cite{wang2017}, which rotates components of the N\'eel and VBS order parameters into each other. Such an emergent symmetry had in fact been observed numerically earlier~\cite{nahum2015b}.
Although having passed a number of consistency checks~\cite{wang2017}, it should be emphasized that this fermion-boson duality, just as most of the other new duality relations in 2+1 dimensions, lacks a formal proof. In particular, it is at present unclear, whether the strong version of the duality holds, implying that the two theories flow to the same renormalization group (RG) fixed point and describe the same infrared physics, or only a weaker version applies, stating that the two theories ``live in the same Hilbert space,'' i.e., they have the same local operators, the same symmetries,  and the same anomalies (if any)~\cite{wang2017}.
Assuming the strong version of the duality, however, implies a number of nontrivial scaling relations between the \cpone\ and \qedgn\ models, as well as new pertinent scaling relations entirely within the \qedgn\ model itself.
These scaling relations allow to test the duality conjecture on a quantitative level and, eventually, to answer the question which version of the duality applies.

Recently, it has been shown that the \qedgn\ model exhibits an infrared stable RG fixed point, the existence of which is a prerequisite for the proposed (strong version of the) duality to hold~\cite{janssen2017}. This can be established within a suitable generalization of the model to noninteger space-time dimension $D$ with $2<D<4$. This theory has an upper critical space-time dimension of $D_\mathrm{c}^+ = 4$, enabling one to compute the critical behavior within a controlled $\epsilon$ expansion in $D=4-\epsilon$ dimensions. Here, we extend the previous one-loop analysis~\cite{janssen2017} to the three-loop order and compute the scaling dimensions of various operators to compare with the predictions from the duality.
Moreover, we establish the previously conjectured~\cite{janssen2017} equivalence of the \qedgn\ model with a gauged fermionic theory in which the boson-mediated scalar interaction is replaced by a corresponding critical four-fermion interaction. This fermionic theory is amenable to a $1/N$ expansion in fixed dimension $2<D<4$, which allows us to demonstrate order by order in a double expansion in both $\epsilon$ and $1/N$ the explicit equivalence of the ultraviolet stable fixed point in this theory with the infrared stable fixed point of the \qedgn\ model.
This puts the asserted equivalence of these gauge theories on the same level as the known ultraviolet-infrared correspondence
between the usual ungauged Gross-Neveu and Gross-Neveu-Yukawa models~\cite{zinnjustin1991, karkkainen1994}.

The rest of the paper is organized as follows:
In Sec.~\ref{sec:model}, we describe the \qedgn\ model, its generalization to noninteger space-time dimension $2<D<4$, as well as the corresponding gauged four-fermion theory. We also briefly comment on the noncompact CP${}^1$ model.
Then, in Sec.~\ref{sec:duality}, we review the corresponding duality conjecture focussing on the consequential relations between the scaling dimensions of different operators.
Sec.~\ref{sec:rg} contains details of the RG scheme and the resulting flow equations. Critical exponents and scaling dimensions are computed in Sec.~\ref{sec:exponents} and estimates for the quantum critical behavior in 2+1~dimensions is presented in Sec.~\ref{sec:resummation}. We discuss these results in light of the conjectured 2+1D boson-fermion duality and give some concluding remarks in the final Sec.~\ref{sec:conclusions}.

\section{Models}\label{sec:model}

Here, we introduce the two relevant U(1)-gauged fermionic models for the dual description of the deconfined quantum critical point, i.e., the \qedgn\ model  and the corresponding gauged four-fermion model.

\subsection{\qedgn\ model}\label{sub:GNY}

The \qedgn\ model is defined in $D=2+1$ Euclidean space-time dimensions by the Lagrangian~\cite{wang2017}
\begin{align}
  \mathcal{L} = \bar{\psi}_i(\slashed{D} + g \phi ) \psi_i + \frac{1}{4}F_{\mu\nu}^2
  +   \frac{1}{2} \phi(r - \partial_\mu^2) \phi + \lambda \phi^4  \label{eq:lagrangian}\,,
\end{align}
with $\psi_i$ and $\bar\psi_i$ being $2N$ flavors of two-component gapless Dirac spinors, $i = 1,\dots,2N$, which interact with each other through the real scalar field $\phi$.
Here, we have abbreviated the covariant derivative $\slashed{D} \equiv (\partial_\mu - i e A_\mu) \sigma_\mu$, with the $2\times 2$ matrices $\sigma_\mu$ serving as a two-dimensional representation of the Clifford algebra, $\{\sigma_\mu, \sigma_\nu\} = 2 \delta_{\mu\nu} \mathds{1}$.
The summation convention over repeated indices is implicitly assumed.
$F_{\mu\nu} = \partial_\mu A_\nu - \partial_\nu A_\mu$ is the field strength tensor of the $\mathrm{U}(1)$ gauge field $A_\mu$, $\mu, \nu \in \{ 0, 1, 2 \}$.
In our calculations, we will also add a gauge-fixing term
\begin{align}
    \mathcal{L}_\text{gf} = - \frac{1}{2\xi}(\partial_\mu A_\mu)^2\,,
\end{align}
to the Lagrangian with gauge fixing
parameter $\xi$, allowing us to check the gauge invariance of our results.
In addition to the $\mathrm{U}(1)$ gauge symmetry, the theory satisfies an $\mathrm{SU}(2N)$ flavor symmetry and a set of discrete symmetries such as parity, charge conjugation, and time reversal.
Under the latter, the scalar field $\phi$ maps to $-\phi$.
The scalar mass parameter~$r$ can be used as a tuning parameter for a symmetry-breaking phase transition at some critical $r_\mathrm{c}$. For $r < r_\mathrm{c}$, $\phi$ acquires a vacuum expectation value, $\langle \phi \rangle \propto \langle \bar\psi_i \psi_i \rangle \neq 0$, signaling the spontaneous breaking of time-reversal symmetry and the dynamical generation of a fermion mass.

A lattice realization of this ordered state is given by the quantum anomalous Hall state with spontaneously generated microscopic currents, a gapped bulk spectrum, and topologically protected chiral edge states~\cite{haldane1988}.
The gauge symmetry as well as the flavor symmetry remain intact across this transition.
In Eq.~\eqref{eq:lagrangian}, the charge $e$, the Yukawa coupling $g$, as well as the bosonic selfinteraction $\lambda$ become simultaneously marginal at the upper critical dimension $D_\mathrm{c}^+ = 4$, suggesting that the critical point may be approached within a standard $\epsilon$ expansion in $D=4-\epsilon$ space-time dimensions.

In order to generalize the theory to arbitrary dimension $2<D<4$, we first combine the $2N$ flavors of two-component spinors into $N$ flavors of four-component spinors~\cite{gehring2015}
\begin{align}
	\Psi_a \equiv
	\begin{pmatrix}
	\psi_a \\
	\psi_{a+N}
	\end{pmatrix}
	\qquad \text{and} \qquad
	\bar\Psi_a \equiv
	\left(
		\bar\psi_a, - \bar\psi_{a+N}
	\right),
\end{align}
with $a = 1, \dots, N$. The Dirac kinetic term then becomes $\bar\Psi_a \partial_\mu \gamma_\mu \Psi_a$ with
$\gamma_\mu = \sigma_z \otimes \sigma_\mu$, where $\sigma_z$ denotes the diagonal $2\times 2$ Pauli matrix.
$\gamma_\mu$ serves as four-dimensional \emph{reducible} representation of the Clifford algebra. The Yukawa interaction reads $\phi \bar\Psi_a \gamma_{35} \Psi_a$ with
$\gamma_{35}  = i \gamma_3 \gamma_5 = \sigma_z \otimes \mathds 1$.
Here, $\gamma_3$ and $\gamma_5$ are the two ``left-over'' gamma matrices, which anticommute with each other as well as with $\gamma_0$, $\gamma_1$, and $\gamma_2$ \cite{herbut2009, gies2010}.
It is important to note that $\gamma_{35}$ squares to one and commutes with the fermion propagator $G_\Psi(p) = -i \delta_{ab} \slashed{p}/p^2$ in $D=3$. The above theory with the full $\mathrm{SU}(2N)$ flavor symmetry therefore has the same loop expansion as a corresponding theory with a smaller symmetry of only $\mathrm{SU}(N) \times \mathrm{SU}(N) \times \mathrm{U}(1)$, in which the Yukawa interaction is replaced by the simple scalar interaction involving only the identity operator
\begin{align}
	\phi \bar\Psi_a \gamma_{35} \Psi_a \mapsto \phi \bar\Psi_a \mathds 1 \Psi_a\,.
\end{align}
This is because in any nonvanishing closed fermion loop, the bilinear operator $\gamma_{35}$ occurs always twice and may thence in all diagrams be replaced by $\mathds 1$ from the outset.
The critical behavior determined by, e.g., the critical exponents $\eta_\phi$ and $\nu$, of these two theories should therefore coincide to all orders in the perturbative expansion. We note, however, that subleading exponents, such as the corrections-to-scaling exponent $\omega$, corresponding to irrelevant operators, might deviate, cf.\ Ref.~\cite{gehring2015}.

A generalization of $\gamma_{35}$ to noninteger dimensions can be obtained by noting that $\gamma_{35} = -\frac{i}{3!} \epsilon_{\mu\nu\rho}\gamma_\mu\gamma_\nu\gamma_\rho = -i \gamma_{[\mu} \gamma_\nu \gamma_{\rho]}$, where the square brackets denote complete antisymmetrization. The last form is a well-defined expression also in noninteger dimension~\cite{dipietro2016}, however, this expression does not preserve the commutation relation between $\gamma_{35}$ and $G_\Psi$ in $D \neq 3$. In $D=4$, e.g., one obtains $\gamma_{35} \to  \gamma_{\mu} \gamma_5$, which neither commutes nor anticommutes with $G_\Psi$.
By contrast, the naive generalization of the $\mathrm{SU}(N) \times \mathrm{SU}(N) \times \mathrm{U}(1)$-symmetric theory, which keeps the simple scalar Yukawa term $\bar\Psi_a \Psi_a$ in all dimensions $2<D<4$, does retain this crucial property of the loop expansion in a trivial way, $[G_\Psi, \mathds 1] = 0$. In this work, we therefore advocate the use of this latter theory to approach the critical behavior of the \qedgn\ model within an $\epsilon$ expansion around the upper critical space-time dimension of $D_\mathrm{c}^+ = 4$.
%

\subsection{Gauged four-fermion theory}\label{sub:GN}
%
In Ref.~\cite{janssen2017}, it was suggested that the infrared stable fixed point in the critical \qedgn\ model can be equivalently understood as an ultraviolet stable fixed point of a gauged four-fermion model with Lagrangian
\begin{align} \label{eq:four-fermion}
  \mathcal{L}_{\text{4-fermi}} = \bar{\Psi}_a \slashed{D} \Psi_a + \frac{1}{4} F_{\mu\nu}^2  +   \frac{g^2}{2r} (\bar\Psi_a \Psi_a)^2\,.
\end{align}
This is reminiscent of the correspondence between the infrared fixed point in the critical Gross-Neveu-Yukawa model and the ultraviolet fixed point of the fermionic Gross-Neveu model~\cite{zinnjustin1991, karkkainen1994}.
The correspondence can be made plausible by means of a Hubbard-Stratonovich transformation, where the quartic interaction is replaced by a Yukawa coupling to an order-parameter field $\phi$, yielding the effective fermion-boson Lagrangian
\begin{align}
  \mathcal{L}_{\text{4-fermi}}' = \bar{\Psi}_a (\slashed{D} + g \phi) \Psi_a + \frac{1}{4} F_{\mu\nu}^2  +  \frac{r}{2} \phi^2\,,
\end{align}
which is equivalent to Eq.~\eqref{eq:lagrangian} up to the presence of the gradient term $\propto \partial^2 \phi^2$, the boson self-interaction $\propto \lambda \phi^4$ and the previously discussed difference in flavor symmetry. Within the $1/N$ expansion, the model~\eqref{eq:four-fermion} has been shown to possess a critical fixed point in all dimensions $2<D<4$~\cite{gracey1992, gracey1993a, gracey1993b}.
Below, we collect additional evidence that the theories defined by Eqs.~\eqref{eq:lagrangian} and \eqref{eq:four-fermion} lie in the same universality class upon a double expansion in both $1/N$ and $\epsilon = 4-D$.
In fact, we carry out this expansion up to linear order in $1/N$ and cubic order in $\epsilon$ and show that the exponents $\eta_\phi$ and $\nu$ precisely coincide order by order in the calculation.

\subsection{Noncompact \cpone\ model}\label{sub:CP1}
%
The bosonic theory that has been proposed~\cite{wang2017} to be dual to the \qedgn\ model is the \cpone\ sigma model describing two complex fields $z_\alpha$, $\alpha = 1,2$. They satisfy the length constraint $\sum_\alpha |z_\alpha|^2 = 1$, and interact via a noncompact $\mathrm{U}(1)$ gauge field $b_\mu$. The Lagrangian can be written as
\begin{align}
	\mathcal{L}_\text{\cpone} = \frac{1}{\kappa} \sum_\alpha \left|(\partial_\mu - i b_\mu) z_\alpha \right|^2,
\end{align}
with the coupling constant $\kappa$, which is marginal for $D=2$ and perturbatively irrelevant for $D>2$.
In order to employ a controlled expansion in fixed $D=2+1$, the above Lagrangian can be generalized to the \cpn\ model by allowing an arbitrary number $\Nb$ of components of $z$, $\alpha = 1,\dots,\Nb$. At large $\Nb$, the model can be shown to possess a quantum critical point at finite $\kappa = \kappa_\mathrm{c}$, separating an ordered phase for $\kappa < \kappa_\mathrm{c}$ from a disordered phase for $\kappa > \kappa_\mathrm{c}$.
The critical exponents have been computed up to the linear order in $1/\Nb$~\cite{halperin1974,irkhin1996, kaul2008}, yielding the correlation-length exponent $\nu$ as
\begin{align} \label{eq:nu-cpn}
	1/\nu_\text{\cpone} & = 1 + \frac{48}{\pi^2 \Nb} + \mathcal O(1/\Nb^2)
\end{align}
and the anomalous dimension $\eta_z$ as
\begin{align}
	\eta_z & = - \frac{4(3+2\xi)}{\pi^2 \Nb} + \mathcal O(1/\Nb^2)\,.
\end{align}
Note that $\eta_z$ depends on the gauge-fixing parameter $\xi$, while $\nu$ is gauge independent.
The case relevant for the deconfined critical point between N\'eel and VBS orders on the square lattice is given by $\Nb = 2$.
The ``spinon fields'' $z = (z_1, z_2)^\mathrm{T}$ then describe the fractionalized quasiparticles at the transition point, and the N\'eel order parameter is $\vec N = z^\dagger \vec \sigma z$. Here, $\vec \sigma$ stands for the three-dimensional vector of Pauli matrices. The anomalous dimension of the N\'eel order parameter is given by $\eta_\text{N\'eel} = 1 + 2\eta_z + 2\eta_{z^\dagger \sigma z}$, where $\eta_{z^\dagger \sigma z}$ denotes the anomalous dimension of the vertex $z^\dagger \vec \sigma z$. To the linear order in $1/\Nb$, it reads
\begin{align}
	\eta_\text{N\'eel} & = 1 - \frac{32}{\pi^2 \Nb} + \mathcal O(1/\Nb^2)\,,
\end{align}
in which the gauge dependence has dropped out, as expected~\cite{kaul2008}.
The coefficient of the leading-order correction $\propto 1/\Nb$ is large, and higher-order calculations are necessary to yield an estimate for the case of $\Nb = 2$.
The VBS order parameter is given by the instanton operator $\mathcal M_b$, which creates a monopole in the gauge field $b_\mu$ with lowest finite topological charge. The scaling dimension of $\mathcal M_b$ has been computed up to next-to-leading order in the $1/\Nb$ expansion~\cite{dyer2015}, yielding the VBS anomalous dimension
\begin{align}
	\eta_\text{VBS} & = 0.249 \Nb - 0.237 + \mathcal O(1/\Nb)\,,
\end{align}
which is well consistent with numerical results for various $\Nb$~\cite{kaul2012}.

\section{Duality conjecture}\label{sec:duality}

\begin{figure}[t]
      \centering
      \includegraphics[width=\columnwidth]{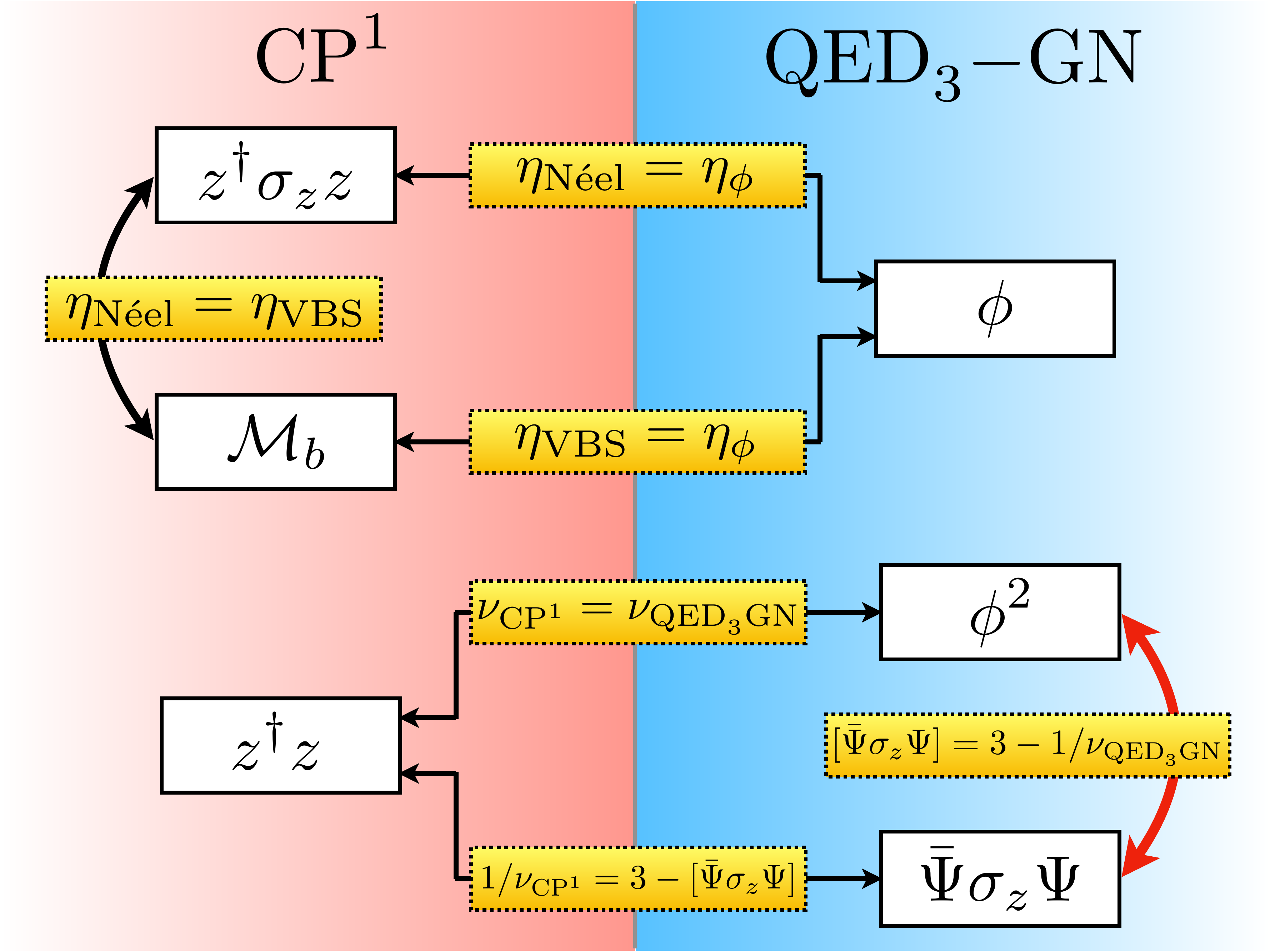}
      \caption{Summary of conjectured duality relations among and within the CP${}^1$ and QED${}_3$-GN models. The duality relation within the $N=1$ QED${}_3$-GN model is emphasized by the thicker red arrow and will be subject to investigation here.}
      \label{fig:duality}
\end{figure}

Let us review the conjectured duality between the \cpone\ model and the $N=1$ \qedgn\ model in 2+1 dimensions~\cite{wang2017}. On the bosonic side, the two real components of the complex VBS order parameter $\mathcal M_b$ and the three components of the N\'eel order parameter $\vec N$ can be combined into a $5$-tuplet,
\begin{align}
	\vec n_\text{\cpone} = (2 \Real \mathcal M_b, 2 \Imag \mathcal M_b, z^\dagger \sigma_x z, z^\dagger \sigma_y z, z^\dagger \sigma_z z)\,.
\end{align}
Here, the first two components of $\vec n_\text{\cpone}$ transform into each other under the global $\mathrm{U}(1)$ symmetry, while the last three components transform as a $3$-vector under the $\mathrm{SU}(2)$ spin symmetry.

On the fermionic side, a monopole in the gauge field $A_\mu$ created by the operator $\mathcal M_A$ induces a zero mode for each of the two Dirac fermions, with one of it filled as a consequence of the Atiyah-Singer index theorem~\cite{nielsen1977}. This allows to construct a $5$-tuplet in the \qedgn\ model with $N=1$ as
\begin{multline}
	\vec n_\text{\qedgn} = \left[
		\Real (\psi_1^\dagger \mathcal M_A),
		-\Imag (\psi_1^\dagger \mathcal M_A),
		\right. \\ \left.
		\Real (\psi_2^\dagger \mathcal M_A),
		\Imag (\psi_2^\dagger \mathcal M_A),
		\phi
	\right].
\end{multline}
The $N=1$ \qedgn\ model has an explicit $\mathrm{SU}(4)$ symmetry in 2+1 dimensions, under which the first four components of $\vec n_\text{\qedgn}$ are rotated into each other.

The proposed strong version of the duality now implies that $\vec n_\text{\cpone}$ and $\vec n_\text{\qedgn}$ are dual to each other when both \cpone\ and \qedgn\ models are tuned to their respective critical points,
\begin{align}
	\vec n_\text{\cpone}(\kappa \to \kappa_\mathrm{c}) \sim \vec n_\text{\qedgn}(r \to r_\mathrm{c})\,.
\end{align}
We emphasize that the duality is expected to hold only precisely in $D=2+1$ dimensions and for two flavors of two-component Dirac spinors on the \qedgn\ side (corresponding to $N=1$ in our notation).
The duality immediately implies an emergent $\mathrm{SO}(5)$ symmetry in both models at criticality, since any component of $\vec n_\text{\cpone}$ and $\vec n_\text{\qedgn}$, respectively can be rotated into each other by applying the symmetry transformation of the respective dual theory.
If the duality holds, it would therefore explain the emergent $\mathrm{SO}(5)$ observed numerically at the deconfined critical point~\cite{nahum2015b}.
It also implies that the scaling dimensions of all components of $\vec n_\text{\cpone}$ and $\vec n_\text{\qedgn}$ coincide. For instance, $[\mathcal M_b] = [z^\dagger \vec \sigma z]$, from which we obtain
\begin{align}
	\eta_\text{VBS} = \eta_\text{N\'eel}\,,
\end{align}
which is consistent with the numerics~\cite{nahum2015a}.
Also, $z^\dagger \sigma_z z \sim \phi$ implies that the anomalous dimensions of the \cpone\ and \qedgn\ models coincide at criticality,
\begin{align}
	\eta_\text{N\'eel} = \eta_\phi\,.
\end{align}
Moreover, from the $\mathrm{SO}(5)$ vectors $\vec n_\text{\cpone}$ and $\vec n_\text{\qedgn}$ we can construct traceless second-rank tensor operators
\begin{align}
	X^{(2)} = \vec n\, \vec n^{\mathrm{T}} - \tfrac{1}{5} (\vec n)^2 \mathds{1}_5\,,
\end{align}
with the duality implying
\begin{align}
	X^{(2)}_\text{\cpone}(\kappa \to \kappa_\mathrm{c}) \sim X^{(2)}_\text{\qedgn}(r \to r_\mathrm{c})\,.
\end{align}
Note that $X^{(2)}_\text{\cpone}$ involves the operators $\mathcal M_b^\dagger \mathcal M_b$ and $\vec N^2$, which tune through the N\'eel-VBS transition, while $X^{(2)}_\text{\qedgn}$ involves $\phi^2$, which tunes through the time-reversal-symmetry-breaking transition in the \qedgn\ model. We therefore have that $[\mathcal M_b^\dagger \mathcal M_b]=[\vec N^2]=[\phi^2]$, and the correlation-length exponents must coincide as a consequence of the duality,
\begin{align}
	\nu_\text{\cpone} = \nu_\text{\qedgn}\,.
\end{align}
Even more interestingly, the fermion bilinear $\bar\psi\sigma_z\psi = \bar\psi_1\psi_1 - \bar\psi_2 \psi_2$ can be understood to also correspond to an element of $X^{(2)}_\text{\qedgn}$~\cite{wang2017}, yielding
\begin{align} \label{eq:bilinear-scaling-relation}
	[\bar\psi\sigma_z\psi]	 = 3 - 1/\nu_\text{\qedgn}
\end{align}
at the critical point of the 2+1D \qedgn\ model. Eq.~\eqref{eq:bilinear-scaling-relation} represents an especially powerful implication of the duality, as it relates the scaling dimensions of different operators of the \emph{same} model to each other, and can thus be fully checked within a standalone \qedgn\ calculation~\cite{wang2017}.
The duality between the \cpone\ and \qedgn\ models and the implications for the critical behaviors are summarized in Fig.~\ref{fig:duality}.
In the following, we will compute the critical behavior of the \qedgn\ model with a particular focus on this scaling relation as a nontrivial check of the conjectured duality.
%

\section{Renormalization group approach}\label{sec:rg}

Here, we first explain the setup for the three-loop renormalization group analysis in $D=4-\epsilon$ space-time dimensions including a few technicalities. Further, we present the full set of beta and gamma renormalization group functions up to three-loop order and explain how to extract the critical fixed point order by order in the $\epsilon$~expansion.

\subsection{Scheme and tool chain}

The bare Lagrangian is defined by Eq.~\eqref{eq:lagrangian} upon
replacing the fields and couplings by their bare counterparts, i.e. $\Psi \mapsto \Psi_0$, $\phi \mapsto \phi_0$, $A_\mu \mapsto A_{0,\mu}$,
$e \mapsto e_0$, $g \mapsto g_0$, $\lambda \mapsto \lambda_0$, $\xi \mapsto \xi_0$.
The renormalized Lagrangian is then introduced as
\begin{align} \label{eq:ren-lagrangian}
  \mathcal{L}' & =
  		Z_\Psi \bar{\Psi}_a \slashed{\partial} \Psi_a -i  Z_{A\bar{\Psi}\Psi} e \mu^{\epsilon/2} \slashed{A}\bar{\Psi}_a\Psi_a  \nonumber \\
        &\quad + \frac{Z_A }{4} F_{\mu\nu}^2 + \frac{Z_\phi}{2} (\partial_\mu \phi)^2 + Z_{\phi^2} \frac{r}{2} \mu^2 \phi^2 \nonumber \\
        &\quad + Z_{\phi\bar{\Psi}\Psi} g \mu^{\epsilon/2} \phi\bar{\Psi}_a\Psi_a + Z_{\phi^4} \lambda \mu^\epsilon \phi^4\,,
\end{align}
where $\mu$ defines the effective energy scale parametrizing the RG flow.
We have defined the wavefunction renormalizations $Z_\Psi$, $Z_\phi$, and $Z_A$, which relate the bare and the
renormalized Lagrangian upon the field rescalings $\Psi_0 = \sqrt{Z_\Psi} \Psi$, $\phi_0 = \sqrt{Z_\phi} \phi$, and $A_{0,\mu} = \sqrt{Z_A}A_\mu$.
The explicit $\mu$ dependencies in the above Lagrangian arise from the introduction of dimensionless couplings.
Demanding that the coefficient in front of the mass term remains invariant under the RG requires $r = r_0 \mu^{-2} Z_\phi Z_{\phi^2}^{-1}$.
The dimensionless couplings are then related to the bare couplings as
\begin{align}
 \alpha  &= e^2_0 \mu^{-\epsilon} Z_\Psi^2 Z_A Z_{A\bar{\Psi}\Psi}^{-2}\,, \\
 y &= g^2_0 \mu^{-\epsilon} Z_\Psi^2 Z_\phi Z_{\phi\bar{\Psi}\Psi}^{-2}\,, \\
 \lambda &= \lambda_0 \mu^{-\epsilon} Z_\phi^2 Z_{\phi^4}^{-1}\,.
\end{align}
In the above equations, we have introduced the dimensionless parameters $\alpha \equiv e^2$ and $y \equiv g^2$, where we have implicitly used the fact that the RG flow must be symmetric under sign changes of $e$ and/or $g$.

We calculate the renormalization factors $Z_x$, $x \in \{ \Psi, \phi, A, \phi^2, \phi\bar{\Psi}\Psi,A\bar{\Psi}\Psi, \phi^4 \}$ up to the three-loop order near the upper critical dimension by employing dimensional regularization and the modified minimal subtraction scheme ($\overline{\text{MS}}$) using a chain of computer algebra tools:
All the Feynman diagrams are generated with the program \texttt{QGRAF}~\cite{nogueira1993}.
At three-loop order, the complete number of diagrams is sizable as shown in Tab.~\ref{tab:diag} for the different types of diagrams corresponding to the respective renormalization group factor $Z_x$.
Further processing of the diagrams is done with the programs \texttt{q2e} and \texttt{exp}~\cite{harlander1998,seidensticker1999} and tracing over matrix structures from the Clifford algebra and tensor reduction of Feynman intergrals is accomplished with \texttt{FORM}~\cite{vermaseren2000,kuipers2013,ruijl2017}.
Before the calculation of Feynman integrals is performed, a reduction to master integrals is carried out via integration-by-parts identities~\cite{czakon2004}.
For the evaluation of beta functions and anomalous dimensions we use a setup where the vertex functions are computed by setting one or two external momenta to zero.
Consequently, the integrals are mapped to massless two-point functions.
These are implemented up to three-loop order in \texttt{MINCER}~\cite{larin1991}.

\begin{table}%
  \caption{Number of diagrams to compute in dimensional regularization to the third loop order.}\label{tab:diag}
  \begin{tabular*}{\linewidth}{@{\extracolsep\fill}crrr}
    \hline\hline
      Diagram type                                                                 & 1-loop       & 2-loop       & 3-loop \\
    \hline
      \raisebox{-.4\height}{\includegraphics[scale=0.8]{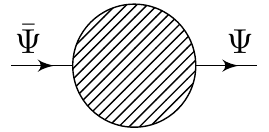}}        & 2            & 13           & 177    \\
      \raisebox{-.4\height}{\includegraphics[scale=0.8]{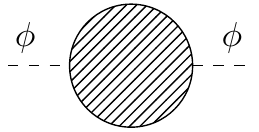}}        & 2            & 9            & 99     \\
      \raisebox{-.4\height}{\includegraphics[scale=0.8]{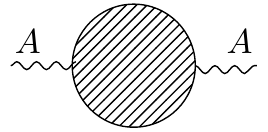}}            & 1            & 6            & 83     \\
      \raisebox{-.4\height}{\includegraphics[scale=0.8]{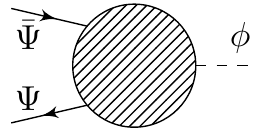}}     & 2            & 38            & 876     \\
      \raisebox{-.4\height}{\includegraphics[scale=0.8]{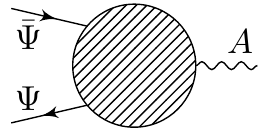}}       & 2            & 37            & 844     \\
      \raisebox{-.4\height}{\includegraphics[scale=0.8]{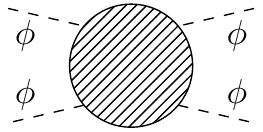}}  & 9            & 153            & 4248     \\
      \hline
      Total                                                                         & 18           & 256          & 6327 \\
    \hline\hline
  \end{tabular*}
\end{table}
%

\subsection{Beta functions}

The beta functions $\beta_x$ for couplings $x \in \{ \alpha, y, \lambda\}$ and the gauge fixing parameter $\xi$ are defined as the logarithmic derivatives with respect to $\mu$ as
\begin{align}
  \beta_x = \frac{\mathrm{d} x}{\mathrm{d} \ln \mu}\,.
\end{align}
We use rescaled couplings $x /(8\pi)^2 \mapsto x$ for $x\in\{\alpha, y, \lambda \}$.
To three-loop order the beta functions can then be written in the form
\begin{align}
    \beta_x = -\epsilon\, x + \beta_x^\text{(1L)} + \beta_x^\text{(2L)} + \beta_x^\text{(3L)}\,,
\end{align}
where we have defined the functions $\beta_x^{\text{($i$L)}}$ to collect the contributions of the
$i$th loop order to the coupling $x~\in~\{ \alpha, y, \lambda \}$.

Up to the three-loop order, the beta function of the gauge coupling $\alpha$ reads
\begin{align}
   \beta_{\alpha}^{(\text{1L})} &= \frac{4}{3} N \alpha ^2\,, \label{eq:betaalpha1}\\
   \beta_{\alpha}^{(\text{2L})} &= 2 N  \alpha^3 - \Nf  y \alpha ^2\,, \\
   \beta_{\alpha}^{(\text{3L})} &= -\frac{\alpha ^2 N }{36} \big[2 \alpha ^2 (22 N+9)-9 (7 N +6) y^2+27 \alpha  y\big].\nonumber
\end{align}
The beta function for the Yukawa coupling $y$ is given by
\begin{align}
   \beta_{y}^{(\text{1L})} &= (3 + 2 \Nf ) y^2 -6 \alpha  y\,, \label{eq:betay1} \\
   \beta_{y}^{(\text{2L})} &= -\left(\frac{9}{8}+6 \Nf \right) y^3 -24 \lambda  y^2 +24 \lambda ^2 y  \nonumber\\
        &\quad +(12 +5 \Nf ) y^2 \alpha  +\frac{1}{6}  (20 \Nf -9) y\alpha ^2\,,  \\
   \beta_{y}^{(\text{3L})} &= \frac{1}{16} y^2 \left[192 \alpha  \lambda +\alpha ^2 \left(-64 \Nf ^2+98 \Nf +327\right) \right. \nonumber\\
        & \left. +48 \lambda ^2 (91-30 \Nf )\right] + \frac{1}{64} \left[2 \Nf  (112 \Nf +67)\nonumber \right. \\
        & \left. -697\right] y^4-\frac{1}{4} y^3 \left[\alpha  (79 \Nf +174)-72 \lambda  (5 \Nf +7)\right]  \nonumber \\
        & +\frac{3}{4} \zeta_3 y \left[-32 \alpha ^3 \Nf +(18 \Nf +19) y^3 \right. \nonumber  \\
        & \left. +4 \alpha  (4 \Nf +3) y^2-12 \alpha ^2 (6 \Nf +7) y\right]-216 \lambda ^3 y  \nonumber \\
        & +\frac{1}{108} \alpha ^3 \left[4 \Nf  (70 \Nf +621)-3483\right] y\,.
\end{align}
$\zeta_z=\zeta(z)$ denotes Riemann's zeta function.
The beta function of the scalar interaction with coupling $\lambda$ reads
\begin{align}
    \beta_{\lambda}^{(\text{1L})} &= 36 \lambda ^2- \Nf  y^2  + 4 \Nf  y \lambda\,, \label{eq:betalambda1} \\
    \beta_{\lambda}^{(\text{2L})} &=   \left(10 \alpha  \lambda -72 \lambda ^2+4 y^2-2 \alpha  y+7 \lambda  y\right) \Nf    y -816 \lambda ^3\,, \nonumber\\
    \beta_{\lambda}^{(\text{3L})} &= 31320 \lambda ^4\!-\!\frac{1}{8} \Nf ^2 y \left(64 \alpha ^2 \lambda\! +\!157 y^3\!-\!868 \lambda  y^2 \right.  \nonumber\\
        & \left. -116 \alpha ^2 y+864 \lambda ^2 y\right) -3 \zeta_3 \left\{\Nf  y \left[-12 \alpha  \lambda  (\alpha +12 \lambda ) \right. \right. \nonumber \\
        & \left. \left. +4 y^3+y^2 (39 \lambda -8 \alpha )+4 y \left(\alpha ^2+19 \alpha  \lambda -81 \lambda ^2\right)\right] \right.\nonumber \\
        &  \left. -6912 \lambda ^4\right\}+\frac{y}{32} \Nf   \left[8 \lambda  \left(6192 \lambda ^2-119 \alpha ^2-1836 \alpha  \lambda \right)  \right. \nonumber \\
        & \left. +5 y^3 -2 y^2 (44 \alpha +4395 \lambda )+4 y \left(131 \alpha ^2+1302 \alpha  \lambda \right. \right. \nonumber \\
        & \left. \left. +4332 \lambda ^2\right)\right]. \label{eq:betalambda3}
\end{align}
Note that the beta functions are gauge independent as expected and we use this as a sanity check on our calculations.
Further, the above expressions fully agree with the QED beta functions~\cite{gorishny1991,gracey2007,zerf2016} in the limit of $y\rightarrow 0$ and $\lambda\rightarrow 0$.
Also, we recover the beta functions of the Ising Gross-Neveu-Yukawa model for $\alpha\rightarrow 0$~\cite{mihaila2017,zerf2017} and the scalar $\phi^4$
theory with Ising symmetry~\cite{kleinert2001}.
Moreover, Eqs.~\eqref{eq:betaalpha1}, \eqref{eq:betay1}, \eqref{eq:betalambda1} are consistent with the one-loop result from Ref.~\cite{janssen2017}.
For completeness, the beta function of the gauge fixing parameter is listed in the appendix.

\subsection{Anomalous dimensions}
%
The field anomalous dimensions $\gamma_x$ are defined by $\gamma_x = \frac{\mathrm{d}\ln Z_x }{\mathrm{d} \ln \mu}$ for $x \in \{ \Psi, A, \phi, \phi^2 \}$ and at three-loop order, they can be expanded as
%
 $
    \gamma_x = \gamma_x^\text{(1L)} + \gamma_x^\text{(2L)} + \gamma_x^\text{(3L)}\,.
  $
%
Explicitly, the boson anomalous dimension is given by
\begin{align}
    \gamma_\phi^\text{(1L)} &= 2 \Nf   y\,,  \\
    \gamma_\phi^\text{(2L)} &= -\frac{5}{2} \Nf  y (y-2 \alpha )+24 \lambda ^2\,,  \\
    \gamma_\phi^\text{(3L)} &= \frac{1}{4} \Nf ^2 y \left(25 y^2-16 \alpha ^2\right)-216 \lambda ^3 \nonumber\\
                & +\frac{1}{32} \Nf  y \left[21 y^2-84 y \alpha -476 \alpha ^2+960 y \lambda \right.  \nonumber \\
                   & \left. -2880 \lambda ^2\right]+\frac{3}{2} \Nf  y \left(y^2-4 y \alpha +12 \alpha ^2\right) \zeta _3 \,.
\end{align}
Note that $\gamma_\phi$ involves no explicit dependence on the gauge-fixing parameter $\xi$, which is consistent with the fact that the scalar-field anomalous dimension determines the exponent in the anomalous power law of the two-point correlator at criticality.

The exponent $\nu$, governing the divergence of the correlation length, can be computed from the renormalization of the mass term, which reads
\begin{align}
    \gamma_{\phi^2}^\text{(1L)} &= -12  \lambda\,  \\
    \gamma_{\phi^2}^\text{(2L)} &= -2 \Nf  y^2+24 \Nf  y \lambda +144 \lambda ^2 \, \\
    \gamma_{\phi^2}^\text{(3L)} &= 36 N y^3-28 N y^2 \alpha -4 N^2 y^2 (4 y-9 \lambda )  \nonumber \\
        &-\frac{33}{2} N y^2 \lambda +153 N y \alpha  \lambda -288 N y \lambda ^2 -6264 \lambda ^3 \nonumber \\
        & -12 \Nf y \left[y^2-3 y (\alpha -5 \lambda )+12 \alpha  \lambda \right] \zeta _3 \,.
\end{align}
The anomalous dimensions for the fermion and the gauge boson are given in the appendix for completeness.
Again, we have checked that these expressions are fully compatible with the known expressions from QED~\cite{gorishny1991,gracey2007}, the Ising Gross-Neveu-Yukawa model~\cite{mihaila2017,zerf2017}, the scalar $\phi^4$
theory with Ising symmetry~\cite{kleinert2001}  and the one-loop results from Ref.~\cite{janssen2017} in the appropriate limits.

We are also interested in the scaling dimension of the flavor-symmetry-breaking bilinear
\begin{align}\label{eq:bilin}
\bar\psi_i (\sigma_z\otimes \mathds{1}_N)_{ij}\psi_j \mapsto \bar\Psi_a (\sigma_z \otimes \mathds{1}_2) \Psi_a \equiv \bar\Psi\sigma_z\Psi\,.
\end{align}
Note that $\sigma_z \otimes \mathds{1}_2 = \gamma_{35}$ in our representation of the 2+1D Clifford algebra. A natural generalization to $D=4-\epsilon$ can be obtained by assuming an even number $N$ of four-component spinors, allowing us to construct an $8\times 8$ operator $\Gamma_{35} = \Gamma_{35}^{-1}$ that commutes with the fermion propagator, $[G_\Psi \otimes \mathds{1}_2, \Gamma_{35}] = 0$. The results for odd $N$ are obtained by analytical continuation. If the \cpone--\qedgn\ duality holds, the scaling dimensions of $[\bar\Psi\sigma_z\Psi]$ and $\phi^2$ coincide for $D = 2 + 1$ and $N = 1$ at the critical fixed point, leading to a nontrivial scaling relation (cf.~Sec.~\ref{sec:duality}).

To calculate the scaling dimension of the bilinear in Eq.~\eqref{eq:bilin}, we introduce an additional term $h \bar\Psi \sigma_z \Psi$, where $h$ serves as an infinitesimal background field that couples linearly to the flavor-symmetry-breaking bilinear.
To leading order in $h$, we obtain
\begin{align}
	\gamma_{\bar\Psi \sigma \Psi}^\text{(1L)} & = 3 \alpha -\frac{3 }{2}y \label{eq:gamma-psisigmapsi-one-loop}\\
	\gamma_{\bar\Psi \sigma \Psi}^\text{(2L)} & = \frac{1}{12} \alpha ^2 (9-20 \Nf )+\frac{7 }{4}\Nf y^2+\frac{9 }{16}y^2 -6 \alpha  y\\
	\gamma_{\bar\Psi \sigma \Psi}^\text{(3L)} & = \frac{1}{216} \alpha ^3 \left(-280 \Nf^2-2484 \Nf+3483\right) +\frac{87 \alpha  y^2}{4}\nonumber\\
  & +\frac{1}{128} \left(176 \Nf^2-604 \Nf+697\right) y^3 +\frac{137}{16} \alpha  \Nf y^2 \nonumber \\
  & -\frac{3}{8} \zeta _3 \left(12 y^2 (\alpha +2 \alpha  \Nf)-32 \alpha ^3 \Nf +19 y^3-84 \alpha ^2 y\right)\nonumber \\
  & +\frac{3}{32} y \left(464 \lambda ^2+\alpha ^2 (80 \Nf-109)\right)-30 \lambda  y^2\,.
\end{align}
We note that Eq.~\eqref{eq:gamma-psisigmapsi-one-loop} agrees with the previous one-loop result from Ref.~\cite{janssen2017}. We have also checked that it is consistent with the QED limit~\cite{dipietro2016,Gracey:2000am} up to three loops.

\subsection{Critical fixed point}

\begin{figure*}[t]
      \includegraphics{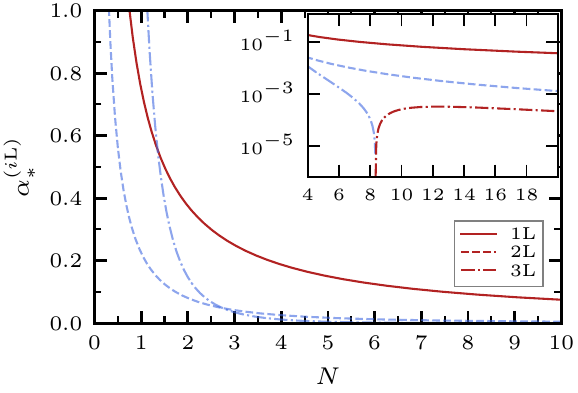}
      \includegraphics{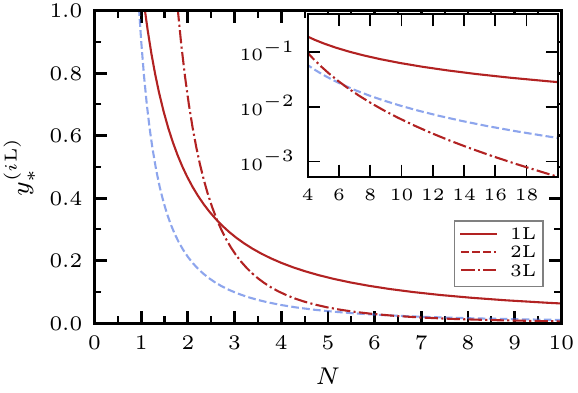}
      \includegraphics{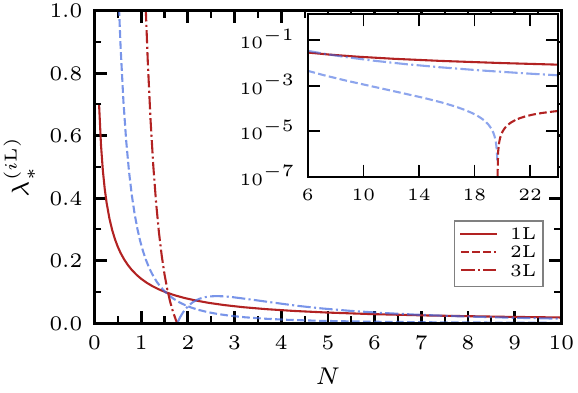}
      \caption{Expansion coefficients $x_*^{(k\mathrm{L})}$ from $k$-loop order of the expansion of the fixed-point values $x_* = \alpha_*$ (left panel), $y_*$ (middle panel), and $\lambda_*$ (right panel) as a function of $N$. The coefficients are large for small $N$, but satisfy a hierarchy  $|x_*^{(3\mathrm{L})}| < |x_*^{(2\mathrm{L})}| < |x_*^{(1\mathrm{L})}|$ at large $N$.
     Darker red (lighter blue) lines denote positive (negative) coefficients.}
      \label{fig: FP Nf dependence}
\end{figure*}

The beta functions allow the determination of the RG fixed points order by order in the $\epsilon$ expansion. To the leading order, we find a unique infrared stable fixed point at
\begin{align}
   (\alpha_*, y_*, \lambda_*) &= \left(\frac{3}{4 \Nf }, \frac{2\Nf +9}{2\Nf (3+2 \Nf )}, \frac{-2\Nf ^2 - 15 \Nf  +s}{72\Nf (3+2 \Nf )} \right) \epsilon \nonumber \\[8pt]
   	& \quad + \mathcal O(\epsilon^2)\,,
\end{align}
where
\begin{align}
s \equiv \sqrt{4\Nf ^4 + 204 \Nf ^3 + 1521 \Nf ^2 + 2916 \Nf }\,,
\end{align}
in agreement with the previous result~\cite{janssen2017}.
The higher-order terms suppressed in the above equation are straightforwardly computed by making use of the beta functions \eqref{eq:betaalpha1}--\eqref{eq:betalambda3}, but we do not display them for general $\Nf$ here for notational simplicity.
At $\Nf =1$ we obtain to the third loop order
\begin{align}
       \alpha_* &= \textstyle \frac{3}{4} \epsilon -\frac{9 }{40}\epsilon ^2 +\frac{430 \sqrt{4645}-101630}{48000}\epsilon ^3+\mathcal O(\epsilon^4) \nonumber \\[5pt]
                &\approx 0.75 \epsilon-0.225 \epsilon ^2-1.50674 \epsilon ^3 + \mathcal O(\epsilon^4) \label{eq:fixed-point-alpha-N1} \\[10pt]
       y_* &= \textstyle \frac{11 }{10} \epsilon+\frac{86 \sqrt{4645}-10741}{5400} \epsilon ^2 \nonumber \\
           & \textstyle \quad +\left(\frac{13263 \zeta_3}{2500}+\frac{271988639}{19440000}-\frac{389515313}{388800 \sqrt{4645}}\right) \epsilon ^3 +\mathcal O (\epsilon^4) \nonumber \\[5pt]
           &\approx 1.1 \epsilon-0.903655 \epsilon ^2+5.66874 \epsilon ^3 +\mathcal O(\epsilon^4) \label{eq:fixed-point-y-N1} \\[10pt]
       \lambda_* &= \textstyle  \frac{\left(\sqrt{4645}-17\right)}{360} \epsilon+\left(\frac{14471}{32400}-\frac{308027}{6480 \sqrt{4645}}\right) \epsilon ^2 \nonumber \\
                 & \textstyle \quad  + \left[\frac{66888 \left(78285599 \sqrt{4645}-6406570729\right) }{50332551120000}\zeta_3 \right. \nonumber \\
                 & \textstyle \quad  \left. +\frac{1814042581409 \sqrt{4645}-132453179835199}{50332551120000}\right] \epsilon ^3 +\mathcal O (\epsilon^4) \nonumber \\[5pt]
                 &\approx 0.142095 \epsilon-0.250827 \epsilon ^2 +1.53577 \epsilon ^3 +\mathcal O(\epsilon^4) \label{eq:fixed-point-lambda-N1}\,.
\end{align}
We note that the three-loop coefficients are large, signaling the divergent behavior of the series at finite $\epsilon \sim \mathcal O(1)$.
For general $\Nf$, the series read
\begin{align}
      x_*(N, \epsilon) = \sum\limits_{k=1}^3 x_{*}^{(k\mathrm{L})}(\Nf) \, \epsilon^k\,,
\end{align}
with $x_* \in \{\alpha_*, y_*, \lambda_*\}$ and expansion coefficients $x_*^{(1\mathrm{L})}$, $x_*^{(2\mathrm{L})}$, and $x_*^{(3\mathrm{L})}$ at one-, two-, and three-loop order, respectively.
The dependence of the expansion coefficients as a function of $N$ is shown in Fig.~\ref{fig: FP Nf dependence}. Note that the higher-loop corrections become small for large $\Nf$ in all three cases.

%

\section{Quantum critical behavior}\label{sec:exponents}

Here, we discuss the critical behavior of the \qedgn\ model, which can be extracted from the $\epsilon$ expansion, i.e. we provide the series expansions for the inverse correlation length exponent, the boson anomalous dimension and the fermion bilinear $\bar\Psi\sigma_z\Psi$ up to order $\mathcal{O}(\epsilon^3)$. For comparison, we also give the corresponding expressions at one-loop order for the four-fermion model and explicitly establish the correspondence between both models in a combined epsilon and $1/N$ expansion.

\subsection{\qedgn\ model}

When the \qedgn\ model is tuned to criticality, the couplings $\alpha$, $y$, and $\lambda$ flow to the infrared stable fixed point and the system becomes scale invariant.
Right at the critical point, the two-point correlation function $G_\phi(p) = \langle \phi(-p) \phi(p) \rangle$ satisfies a power law $G_\phi(p) \propto 1/p^{2-\eta_\phi}$, where the critical exponent $\eta_\phi$ is given by the anomalous dimension at the fixed point,
\begin{align}
	\eta_\phi = \gamma_\phi(\alpha_*,y_*,\lambda_*)\,.
\end{align}
The gauge-field anomalous dimension $\eta_A$ is similarly given by $\eta_A = \gamma_A(\alpha_*,y_*,\lambda_*)$ and governs the power law of the gauge-field propagator $G_A(p) \propto 1/p^{2-\eta_A}$ at the critical point.
Near criticality, the correlation length $\xi_\mathrm{c}$ diverges with exponent $\nu$ as $\xi_\mathrm{c} \propto \left|\delta r_0\right|^{-\nu}$, where $\delta r_0$ measures the distance to the critical point. The correlation-length exponent is obtained from the flow of the dimensionless mass parameter $\beta_r = (-2 + \gamma_\phi - \gamma_{\phi^2}) r$ as
\begin{align}
	 \nu^{-1} = 2 - \eta_\phi + \gamma_{\phi^2}(\alpha_*,y_*,\lambda_*).
\end{align}
We also compute the scaling dimension of the flavor-symmetry-breaking bilinear, which, according to the proposed duality~\cite{wang2017}, is related to $\nu$,
\begin{align}
    [\bar{\Psi}\sigma_z \Psi] = D - 1 - \eta_{\bar{\Psi} \sigma \Psi}
\end{align}
where $\eta_{\bar{\Psi} \sigma \Psi} = \gamma_{\bar{\Psi} \sigma \Psi}(\alpha_*,y_*,\lambda_*)$.

We have computed the scaling dimensions and the critical exponents for all $\Nf$ to the third order in $\epsilon = 4-D$, see appendix. Here, we display the result for the situation of one four-component spinor $\Psi$, which is the case relevant for the duality conjecture. For the scalar-field anomalous dimension $\eta_\phi$ and the correlation-length exponent $\nu$ we find for $\Nf=1$:
\begin{align}
	\eta_\phi &= \textstyle \frac{11}{5}\epsilon +\frac{\left(23 \sqrt{4645}-1768\right)}{900}  \epsilon ^2 \nonumber \\
	&\qquad \textstyle + \left(\frac{91497 }{5000}\zeta_3+\frac{818567}{60000}-\frac{383489}{300 \sqrt{4645}}\right) \epsilon ^3 \\
	&\approx 2.2 \epsilon-0.222725 \epsilon ^2+16.8838 \epsilon ^3 + \mathcal{O}(\epsilon^4) \nonumber
\end{align}
and
\begin{align}
	\nu^{-1} &= 2\textstyle -\frac{\left(\sqrt{4645}+49\right)}{30}  \epsilon+\left(\frac{33311}{54 \sqrt{4645}}-\frac{853}{540}\right) \epsilon^2 \nonumber \\
	&\qquad \textstyle + \left(\frac{39308173}{27000 \sqrt{4645}}-\frac{5931383}{67500}\right) \zeta_3 \epsilon^3 \nonumber \\
	&\qquad \textstyle + \left( \frac{39127961}{2430000}-\frac{1646143919039}{902988000 \sqrt{4645}} \right) \epsilon ^3 \\
	&\approx 2-3.90514 \epsilon+7.47146 \epsilon ^2 -90.5962 \epsilon ^3 + \mathcal{O}(\epsilon^4) \nonumber
\end{align}
We expect hyperscaling to hold at the critical point. The other exponents $\alpha$, $\beta$, $\gamma$, and $\delta$ can hence be obtained from the usual scaling laws~\cite{sachdevbook}.

\begin{figure*}[t]
      \includegraphics{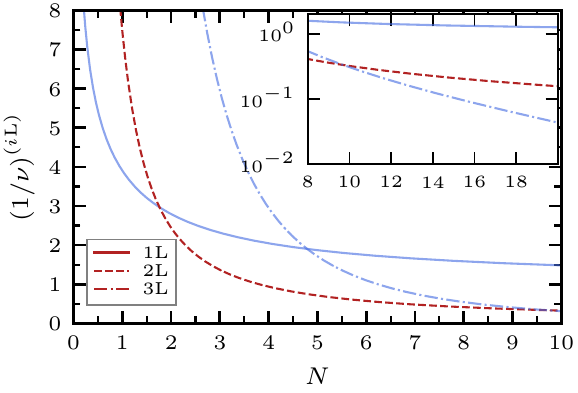}
      \includegraphics{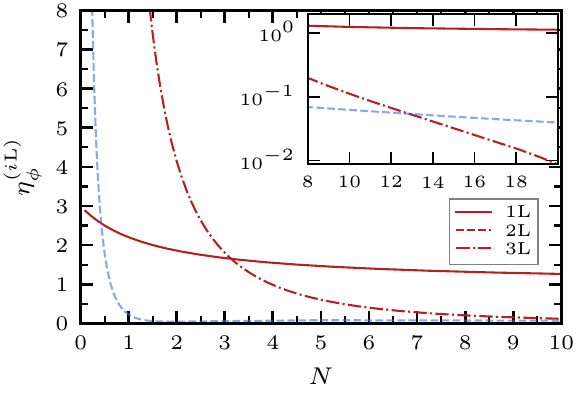}
      \includegraphics{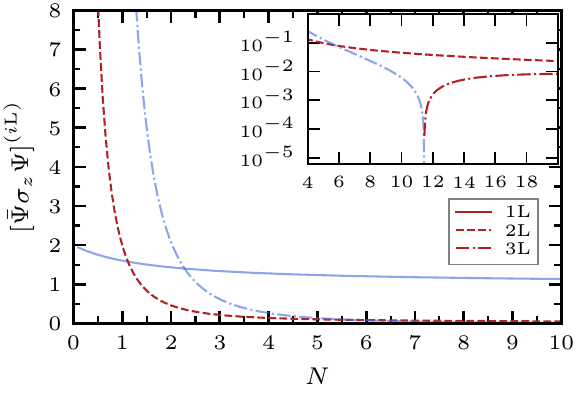}
      \caption{Expansion coefficients of the inverse correlation-length exponent $1/\nu$ (left), the order-parameter anomalous dimension $\eta_\phi$ (middle), and the scaling dimension of the flavor-symmetry-breaking bilinear $[\bar\Psi\sigma_z\Psi]$ (right) as a function of $N$.  Darker red (lighter blue) lines denote positive (negative) coefficients. The higher-loop corrections are again large for $N=1$, but become small for large $N$.}
      \label{fig:expcrit}
\end{figure*}

In order to compare with the $1/N$ expansion of the gauged four-fermion model, it is useful to also compute the exponents $\eta_\phi$ and $1/\nu$ in the limit of large $N$. We find
\begin{align}
	\eta_\phi & = \textstyle
	\left(1+\frac{3}{\Nf}-\frac{9}{2 \Nf^2}\right)\epsilon
	-\left(\frac{1}{\Nf}
	-\frac{39}{8 \Nf^2}\right)\epsilon^2
	\nonumber \\ & \quad \textstyle
	-\left(\frac{3}{4 \Nf}
	-\frac{816 \zeta_3-413}{32 \Nf^2}\right)\epsilon^3
	+\mathcal O(1/N^3,\epsilon^4) \label{eq:eta-largeN}
\end{align}
and
\begin{align}
	\nu^{-1} & = 2
			-\left(1+\frac{6}{N}-\frac{63}{2 N^2}\right) \epsilon
			+\left(\frac{7}{2 N}-\frac{207}{8 N^2}\right) \epsilon^2\notag\\
			&+\left(\frac{1}{N}-\frac{45 \zeta_3}{N^2}+\frac{629}{32 N^2}\right)\epsilon^3
			+\mathcal O(1/N^3,\epsilon^4)\,. \label{eq:nu-largeN}
\end{align}
The scaling dimension of the flavor-symmetry-breaking bilinear reads for $N=1$
\begin{align}
  [\bar{\Psi}\sigma_z \Psi] &= \textstyle 3-\frac{8 }{5} \epsilon  +\frac{\left(43 \sqrt{4645}+646\right) }{1800} \epsilon ^2 \nonumber \\
  	&\quad \textstyle + \left(\frac{51393 }{10000}\zeta_3-\frac{22196749}{3240000}-\frac{37077727}{32400 \sqrt{4645}}\right) \epsilon ^3 \\
	&\approx 3-1.6 \epsilon+1.987 \epsilon ^2 -17.46 \epsilon ^3 + \mathcal{O}(\epsilon^4) \nonumber
\end{align}
whereas in the large-$N$ limit we obtain
\begin{align}
	[\bar{\Psi}\sigma_z \Psi] &= \textstyle 3+\left(-1-\frac{3}{2 \Nf }+\frac{9}{4 \Nf ^2}\right) \epsilon +\left(\frac{1}{2 \Nf } - \frac{15}{16 \Nf ^2} \right) \epsilon^2 \nonumber \\
	&\quad \textstyle +\left(\frac{3}{8 \Nf} +\frac{1-216 \zeta (3)}{64 \Nf ^2} \right)\epsilon ^3+\mathcal{O}\left(1/\Nf ^3,\epsilon ^4\right) \,.
\end{align}
For the gauge-field anomalous dimension $\eta_A$ we find
\begin{align}
	\eta_A = \epsilon + \mathcal{O}(\epsilon^4)\,,
\end{align}
for all $\Nf$, which is consistent with the Ward identity associated with the $\mathrm{U}(1)$ gauge symmetry, requiring $Z_\Psi = Z_{A\bar\Psi\Psi}$ in the renormalized Lagrangian, Eq.~\eqref{eq:ren-lagrangian}.
This result provides another nontrivial crosscheck of our calculations.
The power law of the gauge-field propagator at criticality thus reads $G_A(p) \propto 1/p$ exactly, in agreement with the situation in plain QED$_3$~\cite{pisarski1984, janssen2016, gusynin2016}.

We show the expansion coefficients as a function of $N$ for the inverse correlation-length exponent, the order-parameter anomalous dimension and the scaling dimension of the flavor-symmetry-breaking bilinear in Fig.~\ref{fig:expcrit}. The figure demonstrates that only for large enough $N$ the higher-loop corrections become small.
For small $N$, the series expansions of these other exponents, in contrast to $\eta_A$, exhibit a sizable growth in magnitude, with the three-loop terms for $N=1$ being significantly larger than the leading-order terms. This is in analogy to the notorious situation in multi-loop calculations of the standard bosonic $\mathrm{O}(N)$ models~\cite{kleinert2001}.
The determination of estimates for scaling dimensions of operators in three dimensions therefore requires a suitable resummation scheme.
Due to the lack of knowledge on the large-order behavior of the series, here, we employ simple Pad\'e approximants, see Sec.~\ref{sec:resummation}.

\subsection{Gauged four-fermion model}
%
Here, we compare the exponents $\eta_\phi$ and $\nu$ of the \qedgn\ model with those of the gauged four-fermion model in Eq.~\eqref{eq:four-fermion}. The scaling dimensions of the latter model have been computed before within the $1/N$ expansion for all space-time dimensions $2<D<4$~\cite{gracey1992, gracey1993a, gracey1993b}.
At the critical point, the scalar-field propagator in real space satisfies the power law $G_\phi(x) \propto \left(1/x_\mu^2\right)^{a}$ with exponent~\cite{gracey1993a}
\begin{align}
	a = 1 + \frac{(D-1)\Gamma(D-1)}{2[\Gamma(D/2)]^3 \Gamma(\tfrac{4-D}{2})} \frac{1}{N} + \mathcal O(1/N^2)\,,
\end{align}
where $\Gamma(\,\cdot\,)$ denotes the Gamma function.
From the exponent $a$, we obtain the anomalous dimension $\eta_\phi$ as
\begin{align}
	\eta_\phi & = 4-D + \frac{(D-1)\Gamma(D-1)}{[\Gamma(D/2)]^3\Gamma(\tfrac{4-D}{2})} \frac{1}{N} + \mathcal O(1/N^2) \\
		& = 1 + \frac{16}{\pi^2 N} + \mathcal O(1/N^2)\,, \label{eq:anom-dim-largeN}
\end{align}
where the second line correspond to the physical case of $D=3$.
We note that the $\mathcal O(1/N)$ correction in $\eta_\phi$ is \emph{positive}, indicating an unusually large anomalous dimension $\eta_\phi > 1$, at least as long as $N$ is large. This is in contrast to the situation in the (ungauged) Gross-Neveu model~\cite{vasilev1993, gracey1994}, but consistent with our result in the \qedgn\ model, see Eq.~\eqref{eq:eta-largeN}.

Near, but not right at, the critical point, the scaling of the propagator receives corrections according to
\begin{align}
	G_\phi(x) \propto \frac{1}{\left(x_\mu^2\right)^{a}} \left[1+ c \left(x_\mu^2\right)^{b} + \dots \right],
\end{align}
where $x \equiv \sqrt{x_\mu^2}$, $c$ a constant (with respect to $x$), and the ellipsis denotes higher-order terms that vanish upon approaching the critical point. From the above equation, we can read off the scaling form of the correlation length $\xi_\mathrm{c} \propto |\delta r_0|^{-\nu}$ with $\nu = 1/(2b)$.

Using the result of Ref.~\cite{gracey1993a} for the exponent $b$, we find
\begin{align}
	\nu^{-1}&=D-2-\frac{\Gamma (D+1)}{2\, \Gamma \left(\tfrac{4-D}{2}\right) \Gamma \left(\tfrac{D}{2}\right)^3}\frac{1}{N}+ \mathcal O(1/N^2)\,,\\
		&=1 - \frac{24}{\pi^2 N} + \mathcal O(1/N^2)\,, \label{eq:nu-four-fermion-D3}
\end{align}
where the second line corresponds again to $D=3$.
Expanding the above $1/N$ series for $\eta_\phi$ and $\nu$ further in $\epsilon = 4-D$ allows us to make contact with the exponents of the \qedgn\ model. We find
\begin{align}
	\eta_\phi & =
	\left(1 + \frac{3}{\Nf} \right) \epsilon
	-\frac{\epsilon ^2}{\Nf}
	-\frac{3 \epsilon ^3}{4\Nf}
	+\mathcal O(1/N^2,\epsilon^4)\,.	\label{eq:eta-four-fermion}
\end{align}
and
\begin{align}
	\frac{1}{\nu}=2-\left(1+\frac{6}{N}\right) \epsilon+\frac{7\epsilon^2}{2 N}+\frac{\epsilon ^3}{N}
	 +\mathcal O\left(\frac{1}{N^2},\epsilon^4\right).	\label{eq:nu-four-fermion}
\end{align}
Both Eqs.~\eqref{eq:eta-four-fermion} and \eqref{eq:nu-four-fermion} precisely agree with the corresponding Eqs.~\eqref{eq:eta-largeN} and \eqref{eq:nu-largeN} in the \qedgn\ model. This constitutes yet another useful crosscheck of our calculations. Even more importantly, this result demonstrates the equivalence between the \qedgn\ model and the gauged four-fermion model at criticality, at least in the regime where both $1/N$ and $\epsilon$ expansions are under perturbative control. Put differently, here we have explicitly verified the naive expectation that the presence or absence of the gradient term $\propto \partial^2 \phi^2$ and/or the scalar selfinteraction $\propto \phi^4$ in the \qedgn\ Lagrangian [Eq.~\eqref{eq:lagrangian}] does not change the universality class of the critical system.

\section{Estimates for 2+1D}\label{sec:resummation}

\begin{figure*}[t]
      \includegraphics{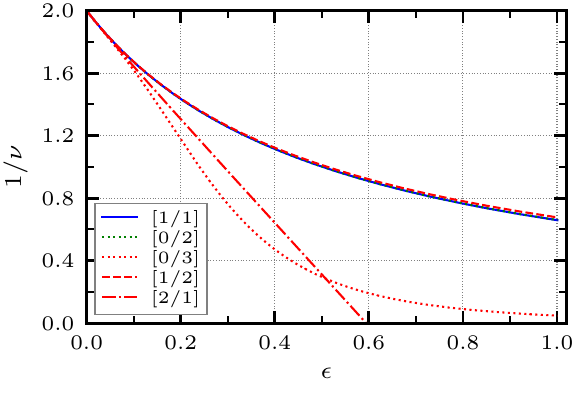}
      \includegraphics{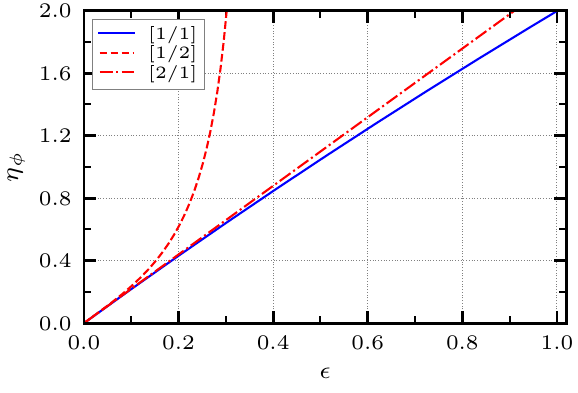}
      \includegraphics{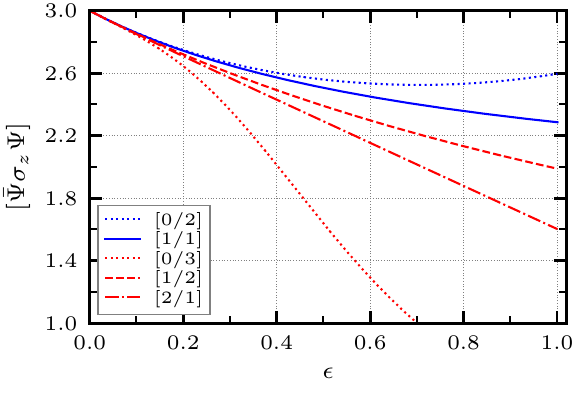}
      \caption{Pad\'e approximants of the inverse correlation-length exponent $1/\nu$ (left panel), the boson anomalous dimension $\eta_\phi$ (mid panel)
      and the bilinear $[\bar{\Psi}\sigma_z\Psi]$ (right panel) at $\Nf =1$. Note that, e.g., the $[1/2]$ approximant of the boson anomalous dimension $\eta_\phi$ has a pole in $\epsilon \in [0,1]$ and has therefore been omitted in Tab.~\ref{tab: Pade approx}.}
      \label{fig:pade}
\end{figure*}

As is usually the case in a perturbative calculation below the upper critical dimension, the resulting series are (at best) asymptotic and diverge upon naively extrapolating to $\epsilon = 1$.
This problem can often be overcome with the help of a suitable resummation scheme, which is a particularly promising approach if the high-order coefficients can be estimated within, e.g., a strong-coupling expansion~\cite{kleinert2001}.
Due to the limited knowledge of the strong-coupling behavior in the present fermionic theories, here we constrain ourselves to a simple Pad\'e approximation.
A comparison with Pad\'e-Borel resummed estimates is deferred to Appendix~\ref{sec: appendix exponents}.

\subsection{Pad\'e approximants}

The Pad\'e approximant for a series expansion $f(\epsilon) = \sum_{k=0}^L f_k \epsilon^k$ truncated at order $L$ is a rational function
\begin{align}
      [m/n] = \frac{a_0 + a_1 \epsilon + \dots + a_m \epsilon^m}{1+ b_1 \epsilon + \dots + b_n \epsilon^n}
\end{align}
with $L = m + n$, and the coefficients $a_0,\dots,a_m$,$b_1,\dots,b_n$ are uniquely given by matching the original series, i.e.,
\begin{align}
      [m/n] - \sum\limits_{k=0}^L f_k \epsilon^k = \mathcal{O}(\epsilon^{L+1}) \,.
\end{align}
The results for the correlation-length exponent $\nu$, the order-parameter anomalous dimension $\eta_\phi$, and the scaling dimension of the flavor-symmetry-breaking bilinear $[\bar\Psi\sigma_z\Psi]$ for different Pad\'e approximants are given for the case of $\Nf=1$ in Tab.~\ref{tab: Pade approx} for $\epsilon = 1$ (corresponding to $D=2+1$) and in Fig.~\ref{fig:pade} as a function of $\epsilon \in [0,1]$.

\begin{table}[tb]
      \caption{Pad\'e-approximated estimates for the inverse correlation-length exponent $1/\nu$, the boson anomalous dimension $\eta_\phi$ and
      the fermion bilinear scaling dimension $[\bar{\Psi}\sigma_z \Psi]$ in the $\Nf = 1$ \qedgn\ model in $D=2+1$ space-time dimensions from two- and three-loop $\epsilon$ expansion.
      Values, for which the approximant has a pole in $\epsilon \in [0,1]$ or is not defined, have been omitted (denoted by ``--''). Estimates from $[0/3]$ strongly deviate from all other Pad\'e approximants and are hence not displayed.
	}
      \label{tab: Pade approx}
\begin{tabular*}{\linewidth}{@{\extracolsep\fill}ccccc}
      \hline\hline
      Order & $[m/n]$		   & $1/\nu$                 & $\eta_\phi$                 & $[\bar{\Psi}\sigma_z \Psi]$ \\
      \hline
      $\epsilon^2$	& $[0/2]$                    & 0.6602                  & --                          & 2.5964          \\
      				& $[1/1]$                    & 0.6595                  & 1.9978                      & 2.2863          \\
      $\epsilon^3$ 	& $[1/2]$                    & 0.6774                  & --  & 1.9894          \\
      			 	& $[2/1]$                    & --    & 2.1971                      & 1.6030          \\
      \hline\hline
\end{tabular*}
\end{table}

A few remarks are in order:
(1) The deviation between the different Pad\'e estimates is not small, in particular for the fermion bilinear. This may point to an inherent strong-coupling nature of the problem, indicated by the large fixed-point values at finite $\epsilon$ for $N=1$, see Eqs.~\eqref{eq:fixed-point-alpha-N1}--\eqref{eq:fixed-point-lambda-N1}. The issue dissolves for larger $N$, for which the loop corrections become small.
(2) The order-parameter anomalous dimension is significantly larger than in the ungauged Gross-Neveu and Gross-Neveu-Yukawa models, for which $\eta_{\phi}^{\text{GN}} \lesssim 1$~\cite{zerf2017,ihrig2018}. Our result that $\eta_{\phi}^{\text{\qedgn}} > 1$ in the \qedgn\ models is consistent with the findings at large $N$, cf.~Eq.~\eqref{eq:anom-dim-largeN}.
(3) The Pad\'e approximation predicts an anomalous dimension of the order of two. Hyperscaling then requires the susceptibility exponent $\gamma$ to (nearly) vanish. An unusual situation would occur if $\eta_\phi$ turned out to be larger than two, leading to a negative $\gamma$ and a vanishing susceptibility at the quantum critical point.

\subsection{Comparison with duality predictions}

We aim to compare our results for the critical behavior of the \qedgn\ model involving two two-component Dirac fermions (corresponding to $N=1$ in our notation) with the predictions from the conjectured duality with the \cpone\ model.
We take the mean values for $1/\nu$, $\eta_\phi$, and $[\bar\Psi\sigma_z\Psi]$, respectively, from the different Pad\'e approximants shown in Tab.~\ref{tab: Pade approx} as our best guesses for the critical exponents in $D=2+1$, and the largest deviation among the different Pad\'e results as an estimate for the order of the confidence interval.
Let us focus on the scaling relation~\eqref{eq:bilinear-scaling-relation} that follows from the assumption of emergent $\mathrm{SO}(5)$ symmetry at the deconfined critical point. This nontrivial relation allows to compare scaling dimensions fully within the \qedgn\ model. For the left-hand-side of the relation, we find
\begin{align} \label{eq:bilinear-scaling-relation-D3}
	[\bar\Psi\sigma_z\Psi] \approx 2.12(50)\,,
\end{align}
which, according to the duality conjecture, should coincide with the right-hand-side
\begin{align} \label{eq:nu-scaling-relation-D3}
	3 - 1/\nu \approx 2.33(1)\,.
\end{align}
We note that in view of the quickly growing series coefficients for $1/\nu$ the spread is likely to be accidentally small. In any case, these estimates are consistent with the duality prediction.

\begin{table}[tb]
\caption{Comparison of critical exponents in 2+1 dimensions for the N\'eel-VBS deconfined critical point described by the noncompact \cpone\ model and the conjectured dual \qedgn\ model for $N=1$.}
\label{tab:comparison}
\begin{tabular*}{\linewidth}{r@{\extracolsep{0pt} }l@{\extracolsep{\fill} }r@{\extracolsep{0pt} }l}
\hline\hline
\multicolumn{2}{c}{\cpone} & \multicolumn{2}{c}{\qedgn\ ($N=1$)} \\
\hline
$\eta_\text{N\'eel} \approx $ & $0.26(3)$ \cite{sandvik2007} & $\eta_\phi \approx $ & $2.1(1)$ [this work] \\
$\approx$ & $0.35(3)$ \cite{melko2008} & $\approx $ & $1.3(9)$ \cite{janssen2017}\\
$\approx$ & $0.30(5)$ \cite{pujari2013}\\
$\approx$ & $0.22$ \cite{bartosch2013}\\
$\approx$ & $0.259(6)$ \cite{nahum2015a} \\
$\eta_\text{VBS} \approx$ & $0.28(8)$ \cite{pujari2013}\\
$\approx$ & $0.25(3)$ \cite{nahum2015a} \\ \\
$3-1/\nu \approx $ & $1.72(5)$ \cite{sandvik2007} & $3-1/\nu \approx $ & $2.33(1)$ [this work] \\
$\approx$ & $1.53(9)$ \cite{melko2008} & $\approx $ & $2.7(4)$ \cite{janssen2017} \\
$\approx$ & $1.15(19)$ \cite{pujari2013} \\
$\approx$ & $1.21$ \cite{bartosch2013} & $[\bar\Psi\sigma_z\Psi] \approx $ & $2.12(50)$ [this work]\\
$\approx$ & $1$ \cite{nahum2015a} & $\approx $ & $1.8(5)$ \cite{janssen2017} \\
$\approx$ & $0.76(4)$ \cite{shao2016} \\
\hline\hline
\end{tabular*}
\end{table}

We should emphasize, however, that our estimates do not agree with the exponents measured in the simulations of the spin systems that are believed to possess a deconfined critical point~\cite{sandvik2007, melko2008, pujari2013, nahum2015a, shao2016}, see Table~\ref{tab:comparison}. In particular, the anomalous dimensions $\eta_\text{N\'eel} \approx \eta_\text{VBS}$ in these bosonic systems (although already being an order of magnitude larger than in the usual Heisenberg or XY universality classes) are significantly below one, while both the $\epsilon$ expansion of the \qedgn\ model in $D=4-\epsilon$ and the $1/N$ expansion of the corresponding four-fermion model in fixed $D=2+1$ find a value for $\eta_\phi^{\text{\qedgn}}$ that is significantly above one. Similarly, the correlation-length exponent $\nu_\text{\cpone}$ in the spin systems appears to be smaller than one, while we find $\nu_\text{\qedgn} \approx 1.50(2) > 1$, in qualitative agreement with the $1/N$ expansion of the four-fermion model [Eq.~\eqref{eq:nu-four-fermion-D3}].
This discrepancy may be interpreted within one of the following three possible scenarios:
\begin{enumerate}[(A)]
\item The strong version of the \cpone--\qedgn\ duality may not hold for the infrared physics and, in this case, the apparent consistency between Eqs.~\eqref{eq:bilinear-scaling-relation-D3} and \eqref{eq:nu-scaling-relation-D3} would be accidental.
\item While the perturbative approach to the \qedgn\ model approximately sustains the duality relation within the model as reflected by Eqs.~\eqref{eq:bilinear-scaling-relation-D3} and \eqref{eq:nu-scaling-relation-D3}, it might not be well-suited to provide reliable absolute estimates for the critical exponents. In that case, non-perturbative approaches, e.g., the functional renormalization group or the conformal bootstrap, could help to check the conjectured duality on the level of critical exponents and scaling relations.
\item The deconfined critical point may really be only a pseudocritical point corresponding to a critical fixed point that has disappeared from the real coupling space as a consequence of a collision and annihilation with another fixed point. 
If indeed existent, any other fixed point would be located outside the perturbative regime for $\epsilon \ll 1$ and can only approach the \qedgn\ fixed point at some finite $\epsilon > 0$.
Such a fixed-point annihilation scenario is known to occur in various gauge theories both in 2+1D and 3+1D~\cite{halperin1974, gies2006, kaplan2009, braun2014, janssen2016, herbut2016, janssen2017a, Roscher:2018ucp}, and has recently been entertained also in the context of deconfined criticality in the spin models~\cite{nahum2015a, wang2017}. In this scenario, $\mathrm{SO}(5)$ would only emerge as an approximate symmetry near a weakly-first-order phase transition with an exponentially large, but finite correlation length $\xi_\mathrm{c}$~\cite{serna2018}. The \cpone--\qedgn\ duality would then only hold at length scales $\ell \lesssim \xi_\mathrm{c}$ in the simulations, and the exponents computed here for the \qedgn\ model would not apply to this pseudocritical regime, but would in fact characterize the nonunitary $\mathrm{SO}(5)$-symmetric fixed point located at complex coupling.
\end{enumerate}

\section{Conclusions}\label{sec:conclusions}

We have determined the critical behavior of the \qedgn\ model within a three-loop $\epsilon$ expansion around the upper critical space-time dimension of $D_\mathrm{c}^+ = 4$.
Within this expansion, the model exhibits a unique infrared stable fixed point corresponding to a continuous phase transition at which a time-reversal symmetry broken fermion mass term is spontaneously generated.
In analogy to the ungauged Gross-Neveu and Gross-Neveu-Yukawa models~\cite{zinnjustin1991}, the infrared fixed point of the \qedgn\ model can be equivalently understood as an ultraviolet fixed point of a corresponding gauged four-fermion theory, the critical behavior of the latter is amenable to a $1/N$ expansion~\cite{gracey1993a}. We have explicitly verified this infrared-ultraviolet correspondence by demonstrating that the critical exponents coincide order by order (up to the order we calculated) within a $(1/N, \epsilon)$ double expansion.

Most interestingly, our estimates for the critical behavior of the 2+1D \qedgn\ universality class for the case of two flavors of two-component Dirac fermions (corresponding to $N=1$ in our notation) are consistent with a nontrivial scaling relation that follows from emergent $\mathrm{SO}(5)$ symmetry implied by the proposed duality between the \qedgn\ and noncompact \cpone\ models~\cite{wang2017}.  If this agreement persists in future calculations that will narrow down our uncertainty interval, it would constitute a strong indication for emergent $\mathrm{SO}(5)$ symmetry at the $N=1$ \qedgn\ fixed point.
Our results, on the other hand, are not compatible with the most recent simulation results for the deconfined critical point in the spin models~\cite{nahum2015b, shao2016}, and we have given a possible interpretation of this discrepancy in terms of the previously proposed~\cite{nahum2015b, wang2017, serna2018} pseudocriticality scenario.

For future work, it would be interesting to study the possible existence of other fixed points that might collide and annihilate with the \qedgn\ fixed point at some space-time dimension between $D=3$ and $D=4-\epsilon$.
As a complementary approach, it would be desirable to test whether the time-reversal-symmetry-breaking transition in the \qedgn\ model is continuous or (weakly) first order, e.g., within a numerical simulation of a suitable lattice model.

\paragraph*{Note added.} After this work was submitted, two related preprints appeared, which confirm our results when considered in the appropriate limits~\cite{zerf2018, gracey2018}.

\paragraph*{Acknowledgments.}
The authors are grateful to J.~A.~Gracey, J.~Maciejko, P.~Marquard, D.~Roscher, and N.~Zerf for very valuable correspondence and comments on the manuscript.
We thank Z.~Y.~Meng and W.~Witczak-Krempa for helpful discussions.
This work has been supported by the Deutsche Forschungsgemeinschaft under projects C04 of SFB1238 (B.I.\ and M.M.S.) and A04 of  SFB1143~(L.J.).

\appendix%
\allowdisplaybreaks[1]%

\section{Further RG functions} \label{sec: appendix gaugefix}

Here, we display the RG functions that have been omitted in the main text.
For the fermion anomalous dimension the corresponding loop contributions read
\begin{align}
    \gamma_\Psi^\text{(1L)} &= \alpha  \xi +\frac{y}{2}\,, \\[3pt]
    \gamma_\Psi^\text{(2L)} &= \frac{1}{4} \alpha ^2 (-4 \Nf -3)-\frac{3}{4}  \Nf  y^2-\frac{y^2}{16}-\alpha  y\,, \\[3pt]
    \gamma_\Psi^\text{(3L)} &= \frac{1}{72} \alpha ^3 \left(40 \Nf ^2+54 \Nf +27\right) \nonumber \\
        & +\frac{1}{128} \left(-48 \Nf ^2+188 \Nf -15\right) y^3 +\frac{5}{16} \alpha  \Nf  y^2 \nonumber \\
        & +\frac{3}{8} \zeta _3 y \left[4 \alpha ^2+4 y (\alpha +2 \alpha  \Nf )+y^2\right] +6 \lambda  y^2\nonumber  \\
        & +\frac{3}{32} y \left[\alpha ^2 (4 \Nf -17)-176 \lambda ^2\right]+\frac{\alpha  y^2}{2}  \,.
\end{align}
We note that only the one-loop term depends on the gauge-fixing parameter $\xi$, cf. also Refs.~\cite{gracey2007, kissler2017}.
The beta function of the gauge fixing parameter follows the form $\beta_\xi = \sum_{k=1}^3 \beta_\xi^{(\text{$k$L})}$ and reads
\begin{align}
    \beta_\xi^{(\text{1L})} &= - \frac{2}{3} \Nf  \alpha \xi \\[3pt]
    \beta_\xi^{(\text{2L})} &=-  \frac{1}{2} \Nf \alpha (2 \alpha -y)  \xi  \\[3pt]
    \beta_\xi^{(\text{3L})} &=  -\frac{1}{72} \Nf \alpha  \xi \bigl[-2 \alpha ^2 (22 \Nf +9) \nonumber \\
      &\quad +9 (7 \Nf +6) y^2-27 \alpha  y\bigr].
\end{align}
These beta functions agree in the limit $y = \lambda \rightarrow 0$ with the QED calculations \cite{gracey2007}.
To explicitly verify the Ward identity associated with the local $\mathrm{U}(1)$ symmetry in our calculations, we also compute the gauge anomalous dimension. We obtain
\begin{align}
    \gamma_A^\text{(1L)} &= \frac{4}{3} \Nf   \alpha\,,  \\[3pt]
    \gamma_A^\text{(2L)} &= (2 \alpha-y)  \Nf  \alpha\,, \\[3pt]
    \gamma_A^\text{(3L)} &= \frac{\alpha}{36} \Nf   \left[9 (6+7 \Nf ) y^2-27 y \alpha -2 (9+22 \Nf ) \alpha ^2 \right].
\end{align}
Gauge invariance requires $\gamma_A = 4-D$ exactly at the critical fixed point~\cite{janssen2017}.

\begin{widetext}
\section{Critical exponents for arbitrary $\Nf$} \label{sec: appendix exponents}
%
We display the critical exponents for general $\Nf$ \cite{supplemental}. It is convenient to abbreviate $s \equiv s(\Nf) = \sqrt{4\Nf^4 + 204 \Nf^3 + 1521 \Nf^2 + 2916 \Nf}$. 
The inverse of the correlation-length exponent then reads
 \begin{align}
            1/\nu & = 2-\frac{10 \Nf^2+39 \Nf+s}{6 \Nf (2 \Nf+3)}\epsilon
            -\frac{1}{108 \Nf (2 \Nf+3)^3 s^2} \left(192 \Nf^8+10672 \Nf^7-96 \Nf^6 s+131232 \Nf^6-8936 \Nf^5 s \right.  \nonumber \\
                  &\left. \hspace{4.5cm} +1196856 \Nf^5-141660 \Nf^4 s+7872660 \Nf^4-835326 \Nf^3 s+27080487 \Nf^3 \right. \nonumber \\
                  &\left. \hspace{4.5cm} -2230713 \Nf^2 s+41504886 \Nf^2  -2985255 \Nf s+21257640 \Nf-2125764 s\right)\epsilon ^2 \nonumber \\[5pt]
            &+\frac{1}{15552 \Nf^3 (2 \Nf+3)^5 s^3} \left(294912 \Nf^{14} \zeta_3-453120 \Nf^{14}+37380096 \Nf^{13} \zeta_3-42546688 \Nf^{13}-147456 \Nf^{12} s \zeta_3 \right. \nonumber \\
                  &\left. \hspace{0.2cm} +917760 \Nf^{12} s+1706655744 \Nf^{12} \zeta_3-1248153088 \Nf^{12}-14929920 \Nf^{11} s \zeta_3+35954979840 \Nf^{11} \zeta_3 \right.  \nonumber \\
                  &\left. \hspace{0.2cm} +66756224 \Nf^{11} s-11880201600 \Nf^{11}+228480 \Nf^{10} s^2  -582100992 \Nf^{10} s \zeta_3+1536213056 \Nf^{10} s \right. \nonumber \\
                  &\left. \hspace{0.2cm} +414397472256 \Nf^{10} \zeta_3+7521785280 \Nf^{10} +12064128 \Nf^9 s^2-11530874880 \Nf^9 s \zeta_3+17570258400 \Nf^9 s \right. \nonumber \\
                  &\left. \hspace{0.2cm} +2913244123392 \Nf^9 \zeta_3+939273904800 \Nf^9-234624 \Nf^8 s^3+152479584 \Nf^8 s^2-118782816768 \Nf^8 s \zeta_3 \right. \nonumber \\
                  &\left. \hspace{0.2cm} +113751808560 \Nf^8 s+13271427877248 \Nf^8 \zeta_3+7692596085888 \Nf^8-4509312 \Nf^7 s^3  +748856448 \Nf^7 s^2 \right. \nonumber \\
                  &\left. \hspace{0.2cm} -689855242752 \Nf^7 s \zeta_3+432301897800 \Nf^7 s+40165847835840 \Nf^7 \zeta_3+30916853322552 \Nf^7 -21600 \Nf^6 s^4\right. \nonumber \\
                  &\left. \hspace{0.2cm} -32667552 \Nf^6 s^3+893187000 \Nf^6 s^2-2383905022848 \Nf^6 s \zeta_3  +943060541292 \Nf^6 s+80251526970144 \Nf^6 \zeta_3\right. \nonumber \\
                  &\left. \hspace{0.2cm} +68128907003406 \Nf^6 -233280 \Nf^5 s^4-103418208 \Nf^5 s^3-5311161576 \Nf^5 s^2-4991366935296 \Nf^5 s \zeta_3 \right. \nonumber \\
                  &\left. \hspace{0.2cm} +1056628909674 \Nf^5 s+101314839073104 \Nf^5 \zeta_3+72296243742564 \Nf^5  -2160 \Nf^4 s^5-882576 \Nf^4 s^4 \right. \nonumber \\
                  &\left. \hspace{0.2cm} -102824640 \Nf^4 s^3-22336611030 \Nf^4 s^2-6205207152672 \Nf^4 s \zeta_3+397375835472 \Nf^4 s+72105783082632 \Nf^4 \zeta_3 \right. \nonumber \\
                  &\left. \hspace{0.2cm} +991212929622 \Nf^4-14904 \Nf^3 s^5-1364688 \Nf^3 s^4+149467356 \Nf^3 s^3  -32684093892 \Nf^3 s^2 \right. \nonumber \\
                  &\left. \hspace{0.2cm} -4220721428112 \Nf^3 s \zeta_3-73196963253 \Nf^3 s +20189256522768 \Nf^3 \zeta_3-73287042502500 \Nf^3-72 \Nf^2 s^6 \right. \nonumber \\
                  &\left. \hspace{0.2cm} -30132 \Nf^2 s^5-739206 \Nf^2 s^4+362560860 \Nf^2 s^3-17065633392 \Nf^2 s^2-1214950653504 \Nf^2 s \zeta_3-18954 \Nf s^5 \right. \nonumber \\
                  &\left. \hspace{0.2cm} +74901294540 \Nf^2 s-2008387814976 \Nf^2 \zeta_3-53054911445616 \Nf^2  -216 \Nf s^6+183347145 \Nf s^3-162 s^6\right)\epsilon ^3\,. \qquad
\end{align}
The order-parameter anomalous dimension is
\begin{align}
        \eta_\phi & = \frac{2 \Nf+9}{2 \Nf+3} \epsilon  -\frac{ 332 \Nf^3+1200 \Nf^2-22 \Nf s+2205 \Nf-93 s+5103}{36 \Nf (2 \Nf+3)^3} \epsilon ^2 \nonumber \\
             &\hspace{0.2cm}-\frac{1}{2592 \Nf^2 (2 \Nf+3)^5 s}\left(12224 \Nf^8+1145920 \Nf^7+56096 \Nf^6 s+24748032 \Nf^6-2115072 \Nf^5 s \zeta_3+1119952 \Nf^5 s \right. \nonumber \\
             &\left. \hspace{0.4cm} +223255440 \Nf^5-16625088 \Nf^4 s \zeta_3+5031576 \Nf^4 s+1031152788 \Nf^4-46422720 \Nf^3 s \zeta_3+2559276 \Nf^3 s \right. \nonumber \\
             &\left. \hspace{0.4cm} +2651693112 \Nf^3-55427328 \Nf^2 s \zeta_3-27824040 \Nf^2 s+3828763404 \Nf^2-25509168 \Nf s \zeta_3-56551446 \Nf s \right. \nonumber \\
             &\left. \hspace{0.4cm} +2593432080 \Nf-2125764 s \zeta_3-34897959 s\right) \epsilon^3\,.
\end{align}
The scaling dimension of the flavor-symmetry-breaking bilinear at the critical point reads
\begin{align}
        [\bar{\Psi}\sigma_z\Psi]  &= 3 -\frac{2 (\Nf+3) }{2 \Nf+3}\epsilon  + \frac{ 284 \Nf^4+552 \Nf^3+2 \Nf^2 s-279 \Nf^2+51 \Nf s+2673 \Nf+162 s}{72 \Nf^2 (2 \Nf+3)^3}\epsilon ^2 \nonumber \\
        &\hspace{0.4cm}-\frac{1}{5184 \Nf^2 (2 \Nf+3)^5 s} \left(2112 \Nf^8+234944 \Nf^7-63264 \Nf^6 s+8115072 \Nf^6+559872 \Nf^5 s \zeta_3-559696 \Nf^5 s \right. \nonumber \\
            &\left. \hspace{0.6cm} +111458160 \Nf^5+1041984 \Nf^4 s \zeta_3+545256 \Nf^4 s+750279564 \Nf^4-8724672 \Nf^3 s \zeta_3+14256756 \Nf^3 s \right. \nonumber \\
            &\left. \hspace{0.6cm} +2710445976 \Nf^3-31492800 \Nf^2 s \zeta_3+42085224 \Nf^2 s+5432779188 \Nf^2-34012224 \Nf s \zeta_3+41551542 \Nf s \right. \nonumber \\
            &\left. \hspace{0.6cm} +6081810804 \Nf-10628820 s \zeta_3+13167927 s+3443737680\right) \epsilon^3 \,.
\end{align}
In the above equations, $\zeta_z \equiv \zeta(z)$ denotes Riemann's zeta function.
\end{widetext}

\section{Pad\'e-Borel resummation} \label{sec: appendix Pade Borel}

\begin{table}[t]
      \caption{Pad\'e-Borel resummed estimates for $D=3$ and $N=1$ of the inverse correlation length $1/\nu$, the boson anomalous dimension $\eta_\phi$ and
      the fermion bilinear's scaling dimension $[\bar{\Psi}\sigma_z \Psi]$.
      Values, for which the approximant has a pole in $\epsilon \in [0,1]$ or is not defined, are omitted (denoted by ``--''). We also do not display the resulting negative value for $1/\nu$ found for $\mathcal{PB}_{[2/1]}$.}
      \label{tab: Pade Borel approx}
\begin{tabular*}{\linewidth}{@{\extracolsep\fill}lrrr}
      \hline\hline
      $\Nf = 1$           & $1/\nu$                 & $\eta_\phi$                 & $[\bar{\Psi}\sigma_z \Psi]$ \\
          \hline
      $\mathcal{PB}_{[0/2]}$     & 0.9321                  & --                          & 2.2003             \\
      $\mathcal{PB}_{[1/1]}$     & 0.3870                  & 2.0056                      & 2.1887             \\
      $\mathcal{PB}_{[0/3]}$     & 0.5965                  & --                          & 1.6922             \\
      $\mathcal{PB}_{[1/2]}$     & --        & --            & 2.1377             \\
      $\mathcal{PB}_{[2/1]}$     & --                 & 2.1958                      & 1.6685             \\
      \hline\hline
\end{tabular*}
\end{table}

Here, we add a discussion on the Pad\'e-Borel resummed estimates for the critical exponents in 2+1D. For this method, we implicitly assume that the
series are Borel summable. The Borel tranform of a series $f(\epsilon)$ is defined as
\begin{align}
      \mathcal{B}^b_f(\epsilon) := \sum\limits_{k=0}^\infty \frac{f_k}{\Gamma(k+b+1)} \epsilon^k\,, \label{eq: Borel transform}
\end{align}
where the parameter $b$ determines the polynomial growth of the overall factorially increasing coefficients
$f_k~\sim~\Gamma(k+b+1)~\approx~k!k^b$. If the series is Borel summable we can analytically continue the Borel
transform to the positive real axis such that the Borel sum
\begin{align}
      \tilde{f}(\epsilon) = \int\limits_{0}^{\infty} \mathrm{d}t\, t^b \mathrm{e}^{-t} \mathcal{B}^b_f(\epsilon t) \label{eq: Borel sum}
\end{align}
is convergent. For a finite-order polynomial the integral in Eq.~\eqref{eq: Borel sum} yields the original series. In order to obtain a resummed series we approximate the Borel transform
by a rational Pad\'e approximant, which is a simple and well-established strategy. Other nonpolynomial approximants, using, e.g., hypergeometric functions \cite{mera2018}, would be equally possible. A more systematic study of different resummation techniques is left for future work.

The 2+1D estimates for the critical exponents $1/\nu$ and $\eta_\phi$ as well as the scaling dimension of the fermion bilinear $[\bar{\Psi}\sigma_z \Psi]$ from the Pad\'e-Borel resummation are listed in Tab.~\ref{tab: Pade Borel approx}. For the latter two, the values and the spread are comparable to the earlier estimates from the simple Pad\'e approximants. We observe, however, that the Pad\'e-Borel resummation scheme yields a larger spread in the prediction of the inverse correlation-length exponent $1/\nu$. Whereas the mean value for $1/\nu$ from Tab.~\ref{tab: Pade Borel approx} is well in agreement with the estimates from Tab.~\ref{tab: Pade approx}, the root-mean-square deviation is now much larger, $1/\nu\approx 0.64(22)$.



\end{document}